\documentclass[longauth]{aa}       

\usepackage{natbib}
\usepackage{graphicx}
\usepackage{txfonts}
\usepackage[colorlinks=true,urlcolor=blue,linkcolor=blue,citecolor=blue]{hyperref}
\usepackage{color}

\DeclareMathAlphabet{\pazocal}{OMS}{zplm}{m}{n}

\newcommand{\PriU}{\pazocal{U}}
\newcommand{\serval}{{\tt serval}}

\begin{document} 
\title{The CARMENES search for exoplanets around M dwarfs}
  \subtitle{Two terrestrial planets orbiting G\,264--012 and one terrestrial planet orbiting Gl\,393\footnote{Tables B.1 and B.2 are only available in electronic form at the CDS via anonymous ftp to cdsarc.u-strasbg.fr (130.79.128.5) or via \url{http://cdsweb.u-strasbg.fr/cgi-bin/qcat?J/A+A/}}
}

\author{
    P.\,J.~Amado\inst{1} 
    \and F.\,F.~Bauer\inst{1}
    \and C.~Rodr\'{\i}guez~L\'{o}pez\inst{1}
    \and E.~Rodr\'{\i}guez\inst{1}
    \and C.~Cardona Guill\'en\inst{2,3}
    \and M. Perger\inst{6,7}
    \and J.\,A.~Caballero\inst{4}
    \and M.~J.~L\'opez-Gonz\'alez\inst{1}
    \and I.~Mu\~noz~Rodr\'{\i}guez\inst{5} 
    \and F.\,J.~Pozuelos\inst{8,25} 
    \and A.~S\'anchez-Rivero\inst{1}
    \and M.~Schlecker\inst{9} 
    \and A.~Quirrenbach\inst{10}
    \and I.~Ribas\inst{6,7}
    \and A.~Reiners\inst{11}
    \and J.~Almenara\inst{12} 
    \and N.~Astudillo-Defru\inst{13} 
    \and M.~Azzaro\inst{14}
    \and V.\,J.\,S.~B\'ejar\inst{2,3}
    \and R.~Bohemann \inst{11} 
    \and X.~Bonfils\inst{12} 
    \and F.~Bouchy\inst{15} 
    \and C.~Cifuentes\inst{4}
    \and M.~Cort\'{e}s-Contreras\inst{4}
    \and X.~Delfosse\inst{12} 
    \and S.~Dreizler\inst{11}
    \and T.~Forveille\inst{12} 
    \and A.\,P.~Hatzes\inst{16}
    \and Th.~Henning\inst{9}
    \and S.\,V.~Jeffers\inst{24}
    \and A.~Kaminski\inst{10}
    \and M.~K\"urster\inst{9}
    \and M.~Lafarga\inst{6,7}
    \and N.~Lodieu\inst{2,3}
    \and C.~Lovis\inst{15} 
    \and M.~Mayor\inst{15} 
    \and D.~Montes\inst{17}
    \and J.\,C.~Morales\inst{6,7}
    \and N.~Morales\inst{1}
    \and F.~Murgas\inst{2,3} 
    \and J.L.~Ortiz\inst{1}
    \and F.~Pepe\inst{15} 
    \and V.~Perdelwitz\inst{18,21} 
    \and D.~Pollaco\inst{19} 
    \and N.\,C.~Santos\inst{20,22} 
    \and P.~Sch{\"o}fer\inst{11}
    \and A.~Schweitzer\inst{21}
    \and N.\,C.~S\'egransan\inst{15} 
    \and Y.~Shan\inst{11}
    \and S.~Stock\inst{10}
    \and L.~Tal-Or\inst{18,11} 
    \and S.~Udry\inst{15} 
    \and M.\,R.~Zapatero~Osorio\inst{23}
    \and M.~Zechmeister\inst{11}
         }

  \institute{
    Instituto de Astrof\'{\i}sica de Andaluc\'{\i}a (IAA-CSIC), Glorieta de la Astronom\'ia s/n, 18008 Granada, Spain 
    \email{pja@iaa.csic.es}
    \and Instituto de Astrof\'{\i}sica de Canarias, V\'{\i}a L\'{a}ctea s/n, 38205 La Laguna, Tenerife, Spain 
    \and Departamento de Astrof\'{\i}sica, Universidad de La Laguna, 38026 La Laguna, Tenerife, Spain 
    \and Centro de Astrobiolog\'{\i}a (CSIC-INTA), ESAC, camino bajo del castillo s/n, 28049 Villanueva de la Ca\~nada, Madrid, Spain 
    \and Departamento de F\'isica Te\'orica y del Cosmos, Universidad de Granada, 18071 Granada, Spain 
    \and Institut de Ci\'{e}ncies de l’Espai (ICE, CSIC), Campus UAB, c/de Can Magrans s/n, 08193 Bellaterra, Barcelona, Spain 
    \and Institut d’Estudis Espacials de Catalunya (IEEC), C/Gran Capit\`a 2-4, 08034 Barcelona, Spain 
    \and Astrobiology Research Unit, University of Li\`ege, All\'ee du 6 ao\^ut, 19, 4000 Li\'ege (Sart-Tilman), Belgium 
    \and Max-Planck-Institut f\"{u}r Astronomie, K\"{o}nigstuhl 17, 69117 Heidelberg, Germany 
    \and Landessternwarte, Zentrum f\"{u}r Astronomie der Universit\"{a}t Heidelberg, K\"{o}nigstuhl 12, 69117 Heidelberg, Germany 
    \and Institut f\"{u}r Astrophysik, Georg-August-Universit\"{a}t, Friedrich-Hund-Platz 1, 37077 G\"{o}ttingen, Germany 
    \and Universit\'e de Grenoble Alpes, CNRS, IPAG, 38000 Grenoble, France 
    \and Departamento de Matem\'atica y F\'isica Aplicadas, Universidad Cat\'olica de la Sant\'isima Concepci\'on, Alonso de Rivera 2850, Concepci\'on, Chile 
    \and Centro Astron\'{o}mico Hispano-Alem\'{a}n (CSIC-Junta de Andaluc\'{\i}a), Observatorio Astron\'{o}mico de Calar Alto, Sierra de los Filabres, 04550 G\'{e}rgal, Almer\'{\i}a, Spain 
    \and Astronomical Observatory of the University of Geneva, 51 ch. Pegasi, CH-1290 Versoix, Switzerland 
    \and Th\"{u}ringer Landessternwarte Tautenburg, Sternwarte 5, 07778 Tautenburg, Germany 
    \and Facultad de Ciencias F\'{\i}sicas, Departamento de F\'{\i}sica de la Tierra y Astrof\'{\i}sica \& IPARCOS-UCM (Instituto de F\'{\i}sica de Part\'{\i}culas y del Cosmos de la UCM), Universidad Complutense de Madrid, 28040 Madrid, Spain 
    \and Department of Physics, Ariel University, Ariel 40700, Israel 
    \and Department of Physics, University of Warwick, Gibbet Hill Road, Coventry CV4 7AL, UK 
    \and Instituto de Astrof\'isica e Ci\^encias do Espa\c{c}o, Universidade do Porto, CAUP, Rua das Estrelas, PT4150-762 Porto, Portugal 
    \and Hamburger Sternwarte, Gojenbergsweg 112, 21029 Hamburg, Germany 
    \and Departamento de F\'isica e Astronomia, Faculdade de Ci\^encias, Universidade do Porto, Rua do Campo Alegre, 4169-007 Porto, Portugal 
    \and Centro de Astrobiolog\'{\i}a (CSIC-INTA), Carretera de Ajalvir km 4, E-28850 Torrej\'on de Ardoz, Madrid, Spain 
    \and Max Planck Institute for Solar System Research, Justus-von-Liebig-weg 3, 37077 G\"ottingen, Germany 
    \and Space sciences, Technologies and Astrophysics Research (STAR) Institute, Universit\' e
  de Li\` ege, 19C All\'ee du 6 Ao\^ ut, B-4000 Li\` ege, Belgium 
}

  \date{Received dd February 2021 / Accepted dd Month 2021}

\abstract{
We report the discovery of two planetary systems, namely G\,264--012, an M4.0 dwarf with two terrestrial planets ($M_{\rm b}\sin{i} = 2.50^{+0.29}_{-0.30}$\,M$_{\oplus}$ and $M_{\rm c}\sin{i} = 3.75^{+0.48}_{-0.47}$\,M$_{\oplus}$), and Gl\,393, a bright M2.0 dwarf with one terrestrial planet ($M_{\rm b}\sin{i} = 1.71 \pm 0.24$\,M$_{\oplus}$).
Although both stars were proposed to belong to young stellar kinematic groups, we estimate their ages to be older than about 700\,Ma. 
The two planets around G\,264--012 were discovered using only radial-velocity (RV) data from the CARMENES exoplanet survey, with estimated orbital periods of $2.30$\,d and $8.05$\,d, respectively.
Photometric monitoring and analysis of activity indicators reveal a third signal present in the RV measurements, at about 100\,d, caused by stellar rotation. 
The planet Gl\,393\,b was discovered in the RV data from the HARPS, CARMENES, and HIRES instruments. 
Its identification was only possible after modelling, with a Gaussian process (GP), the variability produced by the magnetic activity of the star. 
For the earliest observations, this variability produced a forest of peaks in the periodogram of the RVs at around the 34\,d rotation period determined from {\em Kepler} data, which disappeared in the latest epochs. 
After correcting for them with this GP model, a significant signal showed at a period of $7.03$\,d.
No significant signals in any of our spectral activity indicators or contemporaneous photometry were found at any of the planetary periods.
Given the orbital and stellar properties, the equilibrium temperatures of the three planets are all higher than that for Earth. 
Current planet formation theories suggest that these two systems represent a common type of architecture. This is consistent with formation following the core accretion paradigm.
}


\maketitle

\section{Introduction}

Our current understanding on the formation of low-mass planets in close-in orbits around low-mass stars is that they are abundant to at least one per star.
This understanding mainly comes from statistical population analyses of the results provided by ground-based Doppler surveys \citep{Bonf13,Tuomi2014} and NASA's {\em Kepler} mission. 
{\em Kepler} results suggest a planet occurrence of $0.56_{-0.05}^{+0.06}$ Earth-sized planets (1.0--1.5\,R$_{\oplus}$) per M dwarf for periods shorter than 50\,d, going up to $2.5\pm0.2$ planet per star with radii 1--4\,R$_{\oplus}$ and periods shorter than 200\,d \citep{Dressing2015}. 
More recent results, focussing on mid-M dwarfs, determine occurrences of 0.86, 1.36, and 3.07 planets with periods between 0.5 and 10\,d per M3, M4, and M5 dwarf stars, respectively \citep{Hard19}. 

However, despite the large number of these types of planets having been discovered by {\em Kepler} (in both its flavours, the original mission and the K2 extension) and the Doppler surveys, very few have bulk density determinations.
This is the case because a density calculation requires both the radius and mass to be known. Radii can be measured for transiting planets, but only the brightest targets can be followed up on with a Doppler instrument for a mass measurement. On the other hand, Doppler discoveries have low transit probabilities (e.g. 1.5\% in the case of Proxima b; Anglada-Escude+16), making their radii difficult to obtain.

The small number of planets with precise bulk density measurements consist of transiting planets that have mass estimates to a precision better than 20\% (from radial velocity or transit timing variations). From this sample, we learned that planets with masses below 5--6\,M$_{\oplus}$ appear to be rocky with interiors of terrestrial composition \citep{Zeng2016, Lopez-Morales2016}, though they show a large range of possible internal compositions in the 3--5\,M$_{\oplus}$ range.

To be able to understand the influence of different protoplanetary disc conditions in the formation and evolution of planetary systems or the mechanisms of formation of planetary systems challenging current formation theories, such as the Jupiter-like planet GJ~3512~b around a low mass M dwarf \citep{Mora19}, we aim to clarify the differences in the number and type of exoplanets orbiting M dwarfs and their preference to form multi-planetary systems.
To move forward, we need robust statistics for planet occurrence, system architecture, and bulk density, which, in turn, means having a larger number of small planets ($M \lesssim 3\,M_{\oplus}$) with accurate mass and radius measurements.
Current Doppler instruments achieve a precision close to or below 1\,m\,s$^{-1}$ (\citealp[e.g. HARPS]{Mayo03}; \citealp[MAROON-X,][on Gl\,486]{Trif21}; \citealp[ESPRESSO,][on Proxima Centauri]{Suar20}), which is sufficient for detecting these planets orbiting M dwarfs. 

We can improve our statistics by either observing bright M dwarfs to detect transiting planets and follow them up with RVs or the other way around, that is, detecting the planets in blind large RV surveys and next looking for transits. 
The first approach is followed, for instance, by the {\em Transiting Exoplanet Survey Satellite} \citep[{\em TESS},][]{Rick15} mission, a space telescope that was built to find small planets transiting small, bright stars that could be followed up on with RV instruments such as HARPS \citep{Mayo03} or CARMENES \citep{Quir18,Bluh20}.
The second approach is followed by surveys such as CARMENES with the homonymous instrument or RedDots\footnote{\url{https://reddots.space/}} \citep{Jeff20} with HARPS, which are optimised to detect planets around our closest M dwarf neighbours with Doppler measurements, looking for transits next.

In this work, we report the results of a long-term spectroscopic and photometric monitoring campaign of G\,264--012 and Gl\,393, an M4.0V and an M2.0V dwarf star, respectively.
The combined analysis of G\,264--012 confirms the presence of two super-Earths with minimum masses of about 2.5\,M$_\oplus$ and 3.8\,M$_\oplus$ with orbital periods of 2.3\,d and 8.1\,d, respectively.
Around Gl\,393, we detect a super-Earth-type planet with a minimum mass of 1.7\,M$_\oplus$ and a period of 7.0\,d.

The work is organised as follows: Sect.~\ref{sec:star} introduces the main characteristic of the stars; Sect.~\ref{Sec:Data} describes the data sets, the data reduction, and the main characteristics of the instruments used in this study; Sect.~\ref{sec:analysis} presents the analysis and the search for periodic signals; Sect.~\ref{sec:discussion} describes the main characteristics of the planetary system and discusses them; and Sect.~\ref{sec:conclusions} summarises the main results.

\section{The stars}\label{sec:star}

\begin{table*}[]
\caption{\label{tab:parameters} Stellar parameters of G\,264--012 and Gl\,393.}
\centering
\begin{tabular}{lccl}
    \hline
    \hline
    \noalign{\smallskip}
Parameters  & G\,264--012    & Gl\,393   & References$^a$\\
    \noalign{\smallskip}
    \hline
    \noalign{\smallskip}
\multicolumn{4}{c}{\em Identifiers}\\
Name & G\,264--012 & Gl 393 & Gic71 / Gli69 \\
Alt. name(s) & NLTT 52122 & BD+01 2447, Ross 446 & Luy79 / Arg1859, Ros26 \\
Karmn & J21466+668 & J10289+008 & Cab16a \\
    \noalign{\smallskip}
\multicolumn{4}{c}{\em Coordinates, basic photometry, and spectral type}\\
$\alpha$ (J2000) & 21:46:41.25 & 10:28:54.91 & {\em Gaia} EDR3\\
$\delta$ (J2000) & +66:48:12.1 & +00:50:15.9 & {\em Gaia} EDR3\\
$G$ (mag) & $11.7531\pm0.0028$ & $8.6760\pm0.0028$ & {\em Gaia} EDR3\\
$J$ (mag) & $8.837\pm0.021$ & $6.176\pm0.021$ & 2MASS\\
Spectral type & M4.0\,V & M2.0\,V & AF15\\
    \noalign{\smallskip}
\multicolumn{4}{c}{\em Distance and kinematics}\\
$d$ (pc) & $15.986\pm0.010$ & $7.0375\pm0.0025$ & {\em Gaia} EDR3\\
$\mu_{\alpha}\cos{\delta}$ (mas\,a$^{-1}$) & $392.47\pm0.071$ & $-602.992\pm0.024$ & {\em Gaia} EDR3\\
$\mu_{\alpha}$ (mas a$^{-1}$) & $206.812\pm0.062$ & $-731.882\pm0.021$ & {\em Gaia} EDR3\\
$\gamma$ (km\,s$^{-1}$) & $-9.771\pm0.024$ & $+8.005\pm0.023$ & Laf20\\
$U$ (km\,s$^{-1}$) & $-29.238\pm0.018$ & $-7.27\pm0.01$ & This work\\
$V$ (km\,s$^{-1}$) & $-17.089\pm0.023$ & $-27.79\pm0.02$ & This work\\
$W$ (km\,s$^{-1}$) & $-9.164 \pm0.016$ & $-15.45\pm0.02$ & This work\\
    \noalign{\smallskip}
\multicolumn{4}{c}{\em Photospheric parameters}\\
$T_{\rm eff}$ (K) & $3326\pm54$ & $3579\pm51$ & Pas19\\
$\log{g}$ [cgs] & $4.85\pm0.06$ & $4.88\pm0.07$ & Pas19\\
$\rm{[Fe/H]}$ (dex) & $+0.10\pm0.19$ & $-0.09\pm0.16$ & Pas19\\
    \noalign{\smallskip}
\multicolumn{4}{c}{\em Physical parameters}\\
$L$ ($L_\odot$) & $0.01066\pm0.00011$ & $0.02687\pm0.00054$ & Schw19\\
$R$ ($R_{\odot}$) & $0.305\pm0.011$ & $0.426\pm0.013$ & Schw19\\
$M$ ($M_{\odot}$) & $0.297\pm0.024$ & $0.426\pm0.017$ & Schw19\\
    \noalign{\smallskip}
\multicolumn{4}{c}{\em Rotation and activity}\\
$v\sin{i}$ (km\,s$^{-1}$) & $\le 2.0$ & $\le 2.0$ & Rein18\\
$P_{\rm{rot}}$ (d) & $100\pm 6$ & $34.15^{+0.22}_{-0.21}$ & This work \\
pEW(H$\alpha$) (\AA) & $+0.077\pm0.009$ & $-0.005\pm0.007$ & Sch\"o19 \\
$\log{R'_{\rm{HK}}}$ & ... & $-4.91\pm0.06$ & This work \\
$\log{L_{\rm X}}$ (erg\,s$^{-1}$\,cm$^{-2}$) & ... & $26.83^{+0.15}_{-0.11}$ & This work \\
    \noalign{\smallskip}
    \hline
\end{tabular}
\tablebib{
$^a$
2MASS: \cite{Skrutskie2016};
Arg1859: \cite{Argelander1859};
AF15: \cite{AlonsoFloriano2015}; 
Cab16a: \cite{Caba16}
{\em Gaia} EDR3: \cite{gaia20}; 
Gic71: \cite{Gicl71};
Gli69: \cite{Glie69};
Laf20: \cite{Lafa2020};
Luy79: \cite{Luyt79};
New16: \cite{Newton2016};
Pas19: \cite{Passegger2019}; 
Rein18: \cite{Rein18}; 
Ros26: \cite{Ross1926};
Sch\"o19: \cite{Scho19};
Schw19: \cite{Schw19}. 
}
\end{table*}

\object{G\,264--012} and \object{Gl\,393} are two of the approximately 350 M dwarfs monitored during the CARMENES guaranteed time observation programme \citep{Quir14,Rein18}.
Table~\ref{tab:parameters} provides a summary of the basic parameters of the two new exoplanet-host stars with the corresponding references.

G\,264--012 (Karmn~J21466+668) is a nearby, relatively bright M4.0\,V star ($d \sim$ 16\,pc, $J \sim$ 8.8\,mag). It was the subject of investigations on astrometry \citep{Gicl71,Luyt79,LepineShara2005,Dittmann2014,Schneider2016}, low-resolution spectroscopy \citep{Lepine2013,Gaidos2014,AlonsoFloriano2015}, high-resolution spectroscopy \citep{Jeffers2018,Rajpurohit2018,Passegger2018,Passegger2019}, photometric variability \citep{Newton2016}, and speckle imaging \citep{Janson2014}. G\,264--012 was repeatedly catalogued as a potential candidate for exoplanet surveys \citep{LepineGaidos2011,Frith2013,Lepine2013,AlonsoFloriano2015,Rein18}. 

Gl\,393 (Karmn~J10289+008, BD+01~2447) is a closer and brighter M2.0\,V star ($d \sim$ 7.0\,pc, $J \sim$ 6.2\,mag) that was already catalogued in the {\em Bonner Durchmusterung des n\"ordlichen Himmels} \citep{Argelander1859} and by \citet{Ross1926} and, therefore, attracted the attention of many teams worldwide \citep[][among many others]{Wilson1953, Landolt1983, Leggett1992, FischerMarcy1992, Reid1995}. In contrast to G\,264--012, Gl\,393 was extensively monitored in search for exoplanets with RV \citep{IsaacsonFischer2010,Bonf13,Butl17,Gran20}, direct imaging \citep{Masciadri2005,Lafreniere2007,Nielsen2010,Biller2013,Naud2017}, and even radio \citep{Bower2009,Harp2016}.

For the two stars, we collected the photospheric parameters effective temperature $T_{\rm eff}$, surface gravity $\log{g}$, and iron abundance [Fe/H] from \citet{Passegger2019}, which match those from other determinations \citep{Lepine2013, Gaidos2014, Terrien2015, Rajpurohit2018, Passegger2018}. Their metallicities are widely accepted to be roughly solar (see also \citealt{AlonsoFloriano2015} and \citealt{Dittmann2016}). However, the {\em Gaia} DR2 $T_{\rm eff}$ value of G\,264--012, of over 3800\,K \citep{Gaia18a}, is affected by systematics and does not agree with other literature determinations.

The stellar bolometric luminosities $L_*$, radii $R_*$, and masses $M_*$ were taken from \citet{Schw19}.
New, more accurate, $L_*$ values were computed by \citet{Cif20}.
Together with the $T_{\rm eff}$ of \citet{Passegger2019}, the Stefan-Boltzmann law, and the empirical mass-radius relation from M-dwarf eclipsing binaries of \citet{Schw19}, we re-determined $R_*$ and $M_*$ from these new $L_*$, but they differ by less than 2\,\% from the values of \citet{Schw19}, which is well within the uncertainties.
For consistency with previous CARMENES works, we hereafter used the values of \citet{Schw19}, which also agree with previous determinations in the literature \citep[e.g.][]{Dittmann2014,Newton2016}.

To determine the parameters above, we assumed that the two stars are single (i.e. have no brown dwarf or stellar companion at close separations).
For G\,264--012, \citet{Janson2014} ruled out the presence of companions with $\Delta z' \gtrsim$ 7.5\,mag at angular separations greater than about 2\,arcsec (and with $\Delta z' \gtrsim$ 4.0\,mag at $\rho >$ 0.5\,arcsec), which translate into M and L dwarfs at about the substellar boundary at projected physical separations of over approximately 30\,au (8\,au).
For Gl\,393, having been observed with NACO at the VLT, NICI at Gemini-South, and Altair at Gemini-North (see above), the restrictions on the presence of very low-mass companions are even stronger.
For example, in the first high-resolution imaging survey, \citet{Masciadri2005} excluded the presence of objects with $\Delta H \gtrsim$ 13\,mag (9.0\,mag) at angular separations greater than about 2\,arcsec (0.5\,arcsec), which translate into (non-young) L and T dwarfs with masses of a few hundredths of $M_\sun$ at projected physical separations of over approximately 14\,au (3.5\,au).
Even stronger restrictions were imposed by complementary adaptive optics observations in CH$_4$ band at close separations, down to only 2.5\,au, by \citet{Lafreniere2007} and \citet{Biller2013} and with seeing-limited imaging at wide physical separations, of 500--1000\,au, by \citet{Naud2017}.

All previous imaging surveys assumed a very young age for Gl\,393 and, consequently, claimed to reach detection mass limits of a few Jupiter masses.
With pre-{\em Gaia} data, \cite{Zuck2011} and \cite{Malo2013} assigned Gl\,393 to the AB Doradus moving group.
Previously, it was also proposed as a member of the Local Association \citep[also known as Pleiades Moving Group;][and references therein]{Mont2001}, an association that encompasses several moving groups, including AB Doradus \citep{Lope2006}.
The estimated age for the Local Association members ranges from 20\,Ma to 300\,Ma \citep{Lope2009}, while AB\,Dor in particular has a better constrained age similar to that of the Pleiades open cluster \citep[100--125\,Ma;][]{Luhm2005}. 
Therefore, ages in the interval 70--200\,Ma were proposed for Gl\,393 \citep[cf.][]{Masciadri2005,Lafreniere2007,Biller2013,Naud2017}.

\begin{figure*} 
     \includegraphics[width=\linewidth, trim = 0cm 3cm 0cm 3cm, clip]{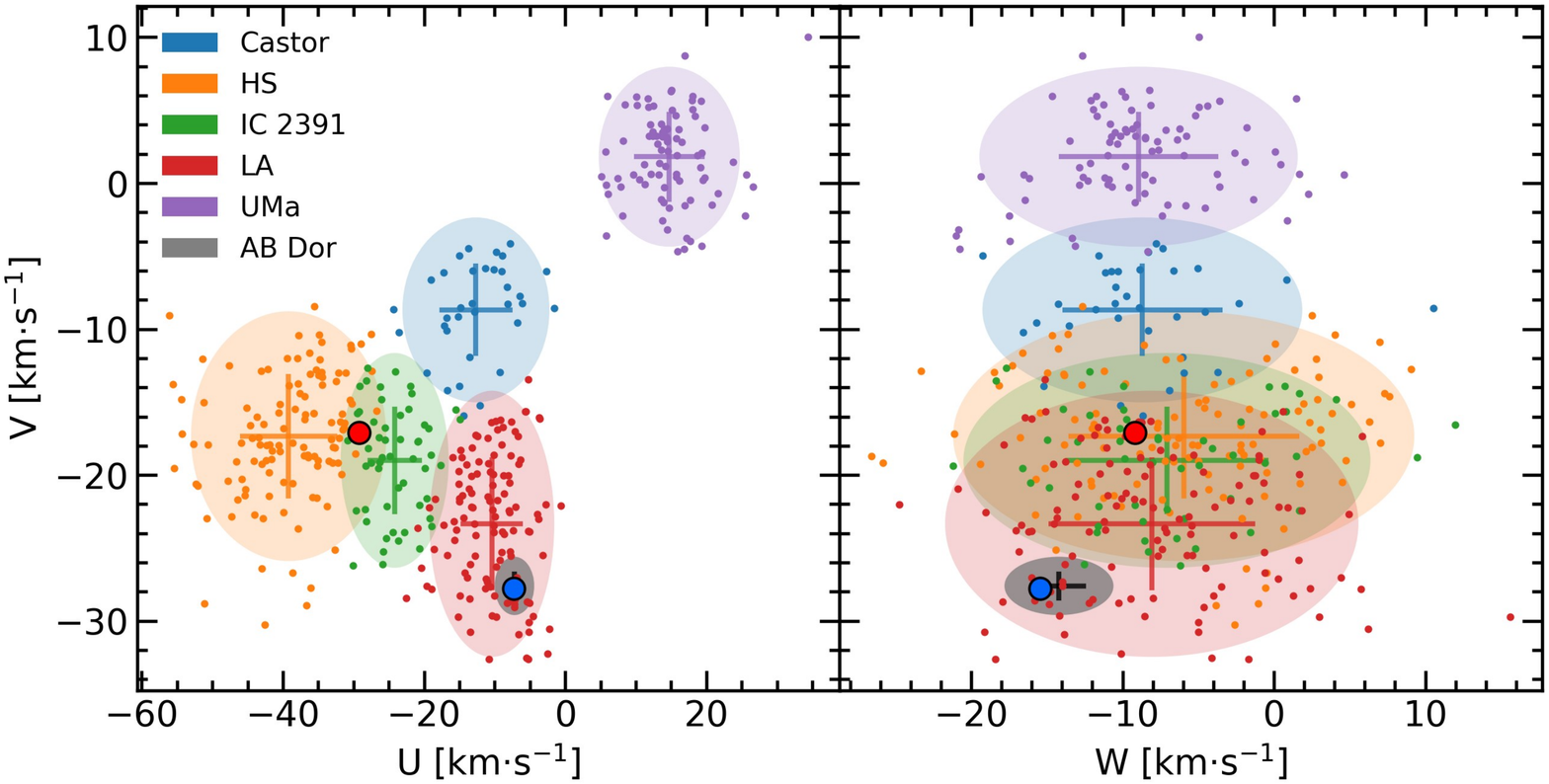} 
     \caption[]{$UVW$ galactocentric space velocities of young moving group members in \citet[][Castor, Hyades Supercluster, IC~2391 Moving Group, Local Association, Ursa Major, AB Doradus]{Mont2001} along with G\,264--012 (in red) and Gl\,393 (in blue). 
     Crosses and ellipses correspond to the $1 \sigma$ and $2 \sigma$ values for each group, respectively. 
     Over-plotted in grey are the typical $UVW$ values and dispersion for AB\,Dor according to \citep{Gagn2018}.}
     \label{fig:UVW} 
\end{figure*} 

For both G\,264--012 and Gl\,393, we recomputed galactocentric space velocities $UVW$ as in \citet{Cort2016}.
We used {\em Gaia} DR2 proper motions and parallaxes and absolute (cross-correlation function) radial velocities from \citet{Lafa2020}.
We compare our new $UVW$ velocities (Table~\ref{tab:parameters}) against those of other members of young stellar kinematic groups from \cite{Mont2001} in  Fig.~\ref{fig:UVW}.
The kinematics of G\,264--012 match those of other members of the IC\,2931 and Hyades superclusters \citep{Egge1958,Egge1991}, with estimated ages of 35--55\,Ma \citep{Egge1995} and 600\,Ma \citep{Egge1958,Egge1998}, respectively. 
As expected, the new kinematics of Gl\,393 is compatible with members of the Local Association and matches the typical $UVW$ for AB\,Doradus \citep{Malo2013,Gagn2018}. 
However, despite both objects being kinematically young, all youth (i.e. activity) indicators available to us suggest otherwise. 
They are slow rotators from the strict upper limit in $v\sin{i}$ of 2\,km\,s$^{-1}$ imposed by \citet{Rein18}.
The rotation period measured from the photometric and spectroscopic analysis in this work (see Sect.~\ref{sec:rotation}), of about 100\,d and 34\,d for G\,264--012 and Gl\,393, respectively, are well above the periods measured for objects with similar spectral type in the Praesepe open cluster \citep{Rebu2017}, whose age \citep[670\,Ma;][]{Doug2019} is similar to the Hyades supercluster.
The H$\alpha$ line in high-quality, high-resolution optical spectra is in very faint emission or even absorption \citep{Jeffers2018,Scho19}.

As described in Sect.~\ref{sec:activity_indicators}, we also measured the intensity of the Ca~{\sc ii} H\&K lines of Gl\,393.
Our mean activity level of $\log{R'_{\rm HK}} = -4.91\pm0.06$ (Table~\ref{tab:parameters}) is in good agreement with values published by \cite{AstudilloDefru2017}, $\log{R'_{\rm HK}} = -5.031\pm0.106$, and~\cite{BoroSaikia2018}, $\log{R'_{\rm HK}} = -4.92$, consistent with the aforementioned rotation periods for M dwarfs, and well below the $\log{R'_{\rm HK}}$ saturation limit at about --4.0
\citep{SuarezMascareno2016,AstudilloDefru2017}.
There are no archival high-resolution spectra covering Ca~{\sc ii} H\&K for G\,264--012.

Gl\,393 was detected in the X-ray range by the {\em ROSAT} all-sky survey \citep{Voge1999}. 
We transformed its count rate and hardness ratio to flux as in \citet{GonAlv14} and, together with {\em Gaia} DR2 parallactic distance, X-ray luminosity.
The measured $\log{L_{\rm X}}$ value, also shown in Table~\ref{tab:parameters}, is much lower than the measured X-ray luminosity of members of moving groups younger than the Hyades supercluster \citep{Stel13}.
G\,264--012 was not detected by {\em ROSAT}.

Both stars were analysed in the exhaustive multi-band photometric analysis of \citet{Cif20}.
While the bluest passband at which G\,264--012 was detected was $B$, Gl\,393 has {\em GALEX} $FUV$ and $NUV$ measurements in the ultraviolet \citep{Mart2005,Bian2014}.
However, its $FUV$ magnitude of $21.57\pm0.42$\,mag is at the survey detection limit and its $NUV-G_{RP}$ colour of $11.542\pm0.086$\,mag puts Gl\,393 in the locus of the least active (and oldest) stars in the colour-spectral type and colour-colour diagrams in Figs.~A.1 and~A.2 of \citet{Cif20}.
Its ultraviolet-optical and infrared colours are also above the expected value for objects younger than the age of the Hyades Supercluster \citep{Find2011}. 

\section{Data}\label{Sec:Data}

\subsection{High-resolution spectroscopic data}\label{sec:hi-res spec}

\begin{table}[]
  \begin{center}
    \caption{Number and quality of used RV observations.}
    \label{tab:RV instruments}
    \begin{tabular}{lccc} 
    \hline
    \hline
    \noalign{\smallskip}
    Instrument & $N_{\rm obs}$ & $\overline{\sigma}_{\rm RV}$ & rms\\
    && [m\,s$^{-1}$] &  [m\,s$^{-1}$] \\
    \noalign{\smallskip}
    \hline
    \noalign{\smallskip}
    \multicolumn{4}{c}{\em G\,264--012}\\
    CARMENES VIS & 159 & 1.7 & 4.3\\
    CARMENES NIR & 146 & 6.5 & 6.5\\
    \noalign{\smallskip}
    \multicolumn{4}{c}{\em Gl\,393}\\
    CARMENES VIS & 84 & 1.5 & 2.3\\
    CARMENES NIR & 74 & 8.0 & 9.1\\
    HARPS & 180 & 1.1 & 2.3\\
    HIRES & 70 & 3.1 & 4.3\\
    \noalign{\smallskip}
    \hline
    \end{tabular}
  \end{center}
\end{table} 

G\,264--012 was monitored only with CARMENES, whereas Gl\,393 was monitored also with another two instruments, namely HARPS and HIRES. 
The RV time series of the two stars and their uncertainties are listed in Tables~\ref{tab:RVs G264} and~\ref{tab:RVs Gl393}. 
Table~\ref{tab:RV instruments} shows a summary of the RVs used in this work and their overall quality, quantified by the mean formal RV uncertainties $\overline{\sigma}_{\rm RV}$ and the root-mean-square (rms) scatter of the RVs around the mean.

\paragraph{CARMENES.} 
We observed G\,264--012 and Gl\,393 as part of the CARMENES\footnote{\url{http://carmenes.caha.es}} guaranteed time observations survey to search for exoplanets around M dwarfs \citep{Rein18}. CARMENES is located at the 3.5\,m telescope of the Calar Alto Observatory in Almer\'{\i}a, Spain, and consists of two independent spectrographs, the visible channel (VIS; 5200--9600\,\AA, $R \approx$ 94\,600) and the near-infrared channel (NIR; 9600--17\,100\,\AA, $R \approx$ 80\,400).  Details on the instrument and its performance were given by, for instance, \cite{Quir14,Quir18}, \cite{Trif18}, and \cite{Bau20}.
We observed G\,264--012 from 19 June 2016 to 17 January 2020, spanning a total of 1310 days. We obtained 171 VIS and 168 NIR spectra with typical exposure times of 1800\,s per spectrum. We rejected spectra with signal-to-noise ratios S/N $< 10$ (VIS: 6; NIR: 7) and spectra without simultaneous Fabry-P\'erot drift measurement (VIS: 6; NIR: 4). As explained by \cite{Bau20}, we further excluded 11 spectra taken with CARMENES NIR prior to the achievement of thermal stability of the NIR channel. Thus, we were left with 159 VIS and 146 NIR spectra suitable for deriving high precision RVs.

Gl\,393 was observed with CARMENES from 10 January 2016 to 17 June 2020, spanning a total of 1621 days, during which we obtained 84 VIS and 74 NIR spectra with typical exposure times of 580\,s per spectrum. We excluded the first 8 NIR spectra in the time series for the same reason as above.

All CARMENES raw spectra were processed using \texttt{caracal} \citep{Caba16,Zech14}, which performs the basic spectral reduction process to obtain calibrated 1D spectra (wavelength calibration using Th-Ne, U-Ne, U-Ar, and Fabry-P\'erot exposures, \citealp{Baue15}; simultaneous drift monitoring using Fabry-P\'erot lamps, \citealp{Scha15}; as well as spectrum extraction, \citealp{Zech14}). The RVs were obtained with \texttt{serval} \citep{Zech18}, a python code that uses a least-square-matching approach between each individual spectrum and a high S/N template, as described by \cite{Guillem_TERRA}. They were corrected for barycentric motion, secular perspective acceleration, instrumental drift, and nightly zero-point variations \citep{Trif18,Trif20,Tal19}. 

\paragraph{HARPS.}
Gl\,393 was also observed, from December 2003 to June 2017, with the High Accuracy Radial velocity Planet Searcher \cite[HARPS,][]{Mayo03}, a precise optical \'echelle spectrograph with a spectral resolution of $R \approx$ 110\,000 installed at the ESO 3.6\,m telescope at La Silla Observatory, Chile. 
We used 180 public HARPS RV datapoints corrected by \cite{Trif20}, who ($i$) computed RVs with \texttt{serval} that are, on average, more precise than the standard-pipeline DRS RVs by a few percent, ($ii$) accounted for nightly zero-point offsets that correct for HARPS systematic effects, 
and ($iii$) took into account the discontinuous RV jump caused by the fibre exchange in 2015.

\paragraph{HIRES.} 
The High Resolution Echelle Spectrometer \citep[HIRES;][]{Vogt94} is installed at the Keck I telescope in Hawai'i, USA. 
HIRES uses the iodine cell technique \citep{Butl96} to obtain RV measurements with a typical precision of a few m\,s$^{-1}$. 
We used 70 archival HIRES datapoints for Gl\,393 to extend our time baseline  from February 1997 to February 2014, which helped us with the analysis of possible long-period signals and to confirm the planetary signal found in the HARPS and CARMENES data sets. 
We used the HIRES data corrected by \cite{Tal19}, who accounted for nightly zero-point offsets and an instrumental jump in 2004, which is an improvement over the original data reduction by \cite{Butl17}.

\subsection{Photometric monitoring}\label{sec:Phot}

\begin{table*}
    \caption{Photometric data sets for G\,264--012 and Gl\,393$^a$.}
    \label{table:photometry}
    \centering
    \begin{tabular}{l ccc cc cc}
    \hline
    \hline
    \noalign{\smallskip}
Data set     &   Season   &  Filter  &   $\Delta t$ & $N_{\rm obs}$ &  $N_{\rm nights}$  &  rms  & $P_{\rm rot}$  \\
             &            &          &      (d)    & &          & (mmag) &      (d) \\
    \noalign{\smallskip}
    \hline
    \noalign{\smallskip}
    \multicolumn{8}{c}{\em G\,264--012}\\
MEarth-1     & 2008--2009   & RG715         &   211 & 225 &   19  &    5.9 &        ...$^b$        \\
MEarth-2     &    2011     & $I_{715-895}$ &    89 & 252 &   11  &    7.7 &        ...$^b$        \\
MEarth-3     & 2011--2015   & RG715         &  1480 & 462 &   96  &    6.3 &        ...$^b$        \\
ASAS-SN      & 2015--2018   &  $V$            &  1316 & 222$^c$ &  222  &   17.0 & 92.8 $\pm$ 0.8 \\
T90/OSN $V$        &    2019     &  $V$            &   265 &1413 &   78  &    5.4 & 99.7 $\pm$ 1.0 \\
T90/OSN $R$        &    2019     &  $R$            &   265 &1389 &   78  &    5.3 &107.5 $\pm$ 1.2 \\
{\em TESS}         &   2019--2020 & $T_{600-1000}$          &   244 &50363 &  78  &    1.5 &        ...$^b$     \\
    \hline
    \noalign{\smallskip}
    \multicolumn{8}{c}{\em Gl\,393}\\
ASAS         & 2001--2009   &  $V$            &  2969 & 358 &  358  &   14.0 &        ...$^b$  \\ 
SuperWASP    & 2008--2014   & $R_{400-700}$          &  2321 &74021&  445  &    7.6 &        ...$^b$ \\
{\em Kepler}/K2    &   2017      & $K_{493-897}$        & 78    &3393 &   78  &    0.6 & 34.0 $\pm$ 0.1 \\
LCOGT 1.0\,m     &   2020      &  $V$            &    72 & 181 &   38  &    7.9 &        ...$^b$    \\ 
LCOGT 0.40\,m  &   2020      &  $V$            &    40 & 301 &   31  &   41.4 &        ...$^b$       \\ 
    \noalign{\smallskip}
    \hline
    \end{tabular}
    \tablefoot{$^a$ The table header shows the data set identifier, the season, the filter used, the time span ($\Delta t$) of the observations, the number of individual observations (N$_{\rm obs}$), the number of nights (N$_{\rm nights}$), the rms of the residuals after removing the strongest signal in each data set, and the rotation period (P$_{\rm rot}$) estimation with formal error bars. 
    $^b$ Since no significant signal is found, tabulated rms is determined on the original time series.
    $^c$ Observations binned to one data point per night.
    }
\end{table*}

We complemented our spectroscopic data with public and proprietary photometric light curves to study the rotation period of both stars. Variability caused by stellar activity, such as spots on the photosphere of M dwarfs, can lead to RV variations not attributable to planets \citep[][and references therein]{Baroch2020}.

We used archival ground-based photometry from ASAS, MEarth, and ASAS-SN, complemented with space-mission photometry from {\em Kepler} K2 and {\em TESS}. We also used unpublished data of the SuperWASP survey.  In addition, we carried out our own photometric monitoring observations with the T90 telescope of the Sierra Nevada Observatory (OSN) and several telescopes of the Las Cumbres Observatory Global Telescope (LCOGT) network. These data sets are summarised in Table~\ref{table:photometry}, while the corresponding observing facilities and acquired data are described below.

\subsubsection{Our observations}

\paragraph{T90 at OSN.} 
The T90 telescope at OSN\footnote{\url{https://www.osn.iaa.csic.es/en}} in Granada (Spain) was used to obtain ground-based photometric observations of G\,264--012 quasi-simultaneous with the spectroscopic ones. T90 is a 0.9\,m Ritchey-Chr\'etien telescope equipped with a CCD camera VersArray 2k$\times$2k with a resulting field of view (FOV) of 13.2$\times$13.2\,arcmin$^2$. 
The camera is based on a high quantum efficiency back-illuminated CCD chip, type 
Marconi-EEV CCD42-4, with optimised response in the ultraviolet. Our set of observations, collected in Johnson $V$ and $R$ filters, consists of 78 epochs obtained between April 2019 and January 2020.  Each epoch typically consisted of 20 observations of 80\,s per night and per filter. 

The resulting light curves were obtained via synthetic aperture photometry. Each CCD frame was corrected in a standard way for bias and flat field.
Different aperture sizes were tested in order to choose the best one for our observations.
A number of nearby and relatively bright stars within the frames were selected as check stars in order to choose the best ones to be used as reference stars \citep[for further details][see]{Perger2019}.
The mean formal uncertainty, $\sigma$, of the data points of each night is of about 3.4 and 2.7\,mmag in $V$ and $R$ bands, respectively.

\paragraph{LCOGT.} 
Las Cumbres Observatory Global Telescope network\footnote{\url{https://lco.global}} \citep[LCOGT,][]{Brown2013} is a worldwide robotic telescope network that includes 0.4, 1.0, and 2.0\,m telescopes. 
In particular, the 0.4\,m telescopes are equipped with 2k$\times$3k CCDs with a 19$\times$29\,arcmin$^2$ FOV, while the 1.0\,m telescopes are equipped with 4k$\times$4k 
CCDs with a 27$\times$27\,arcmin$^2$ FOV.
We used one 0.4\,m telescope located at Observatorio del Teide in Tenerife (Spain) and three 1.0\,m telescopes: two located at Siding Springs Observatory (Australia) and one at McDonald Observatory (Texas, USA).
We acquired data in the $V$ filter only for Gl\,393 in the period April--June 2020.

\subsubsection{Photometric monitoring surveys}

\begin{figure} 
    \includegraphics[width=0.5\textwidth]{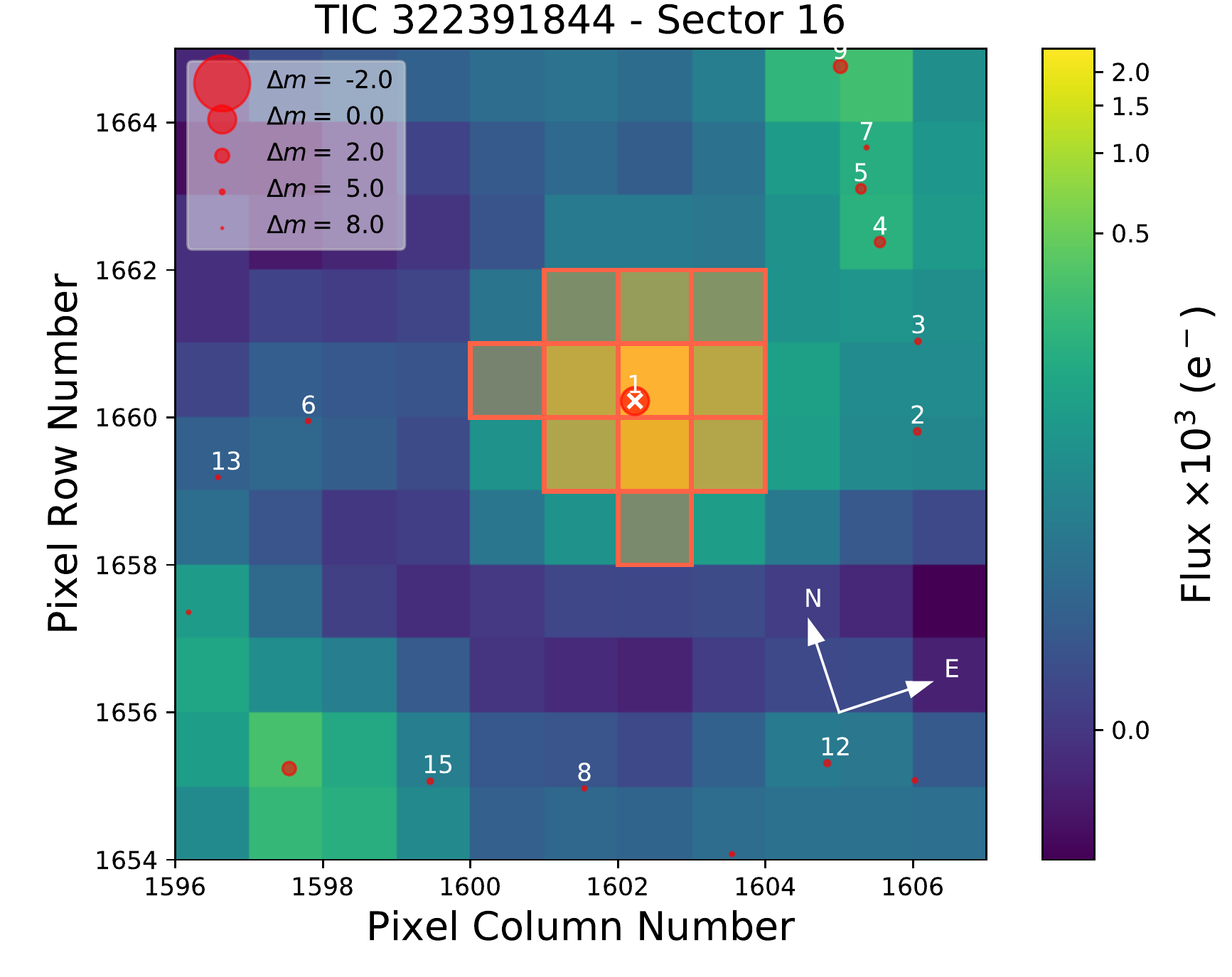} 
    \includegraphics[width=0.5\textwidth]{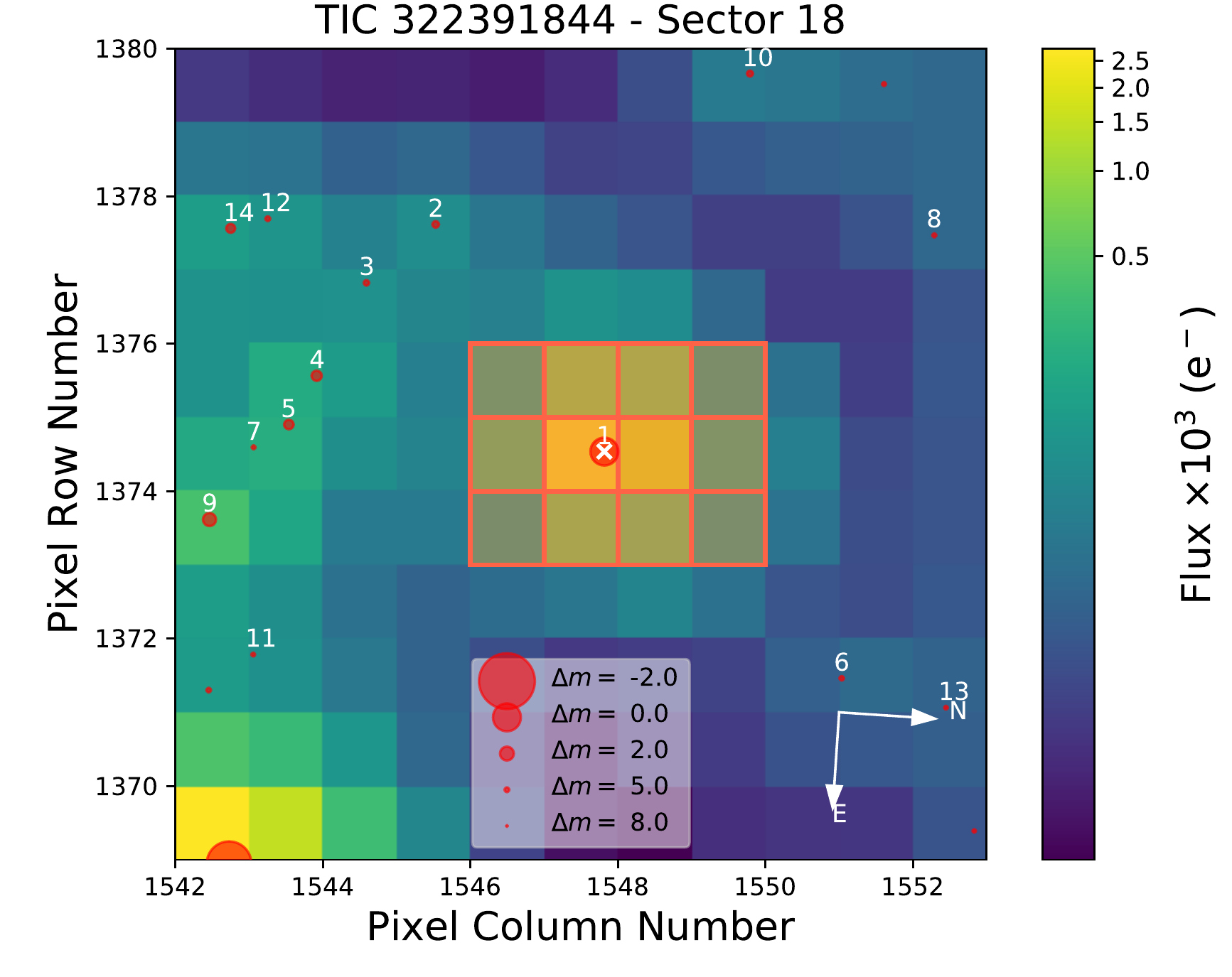}  
    \includegraphics[width=0.5\textwidth]{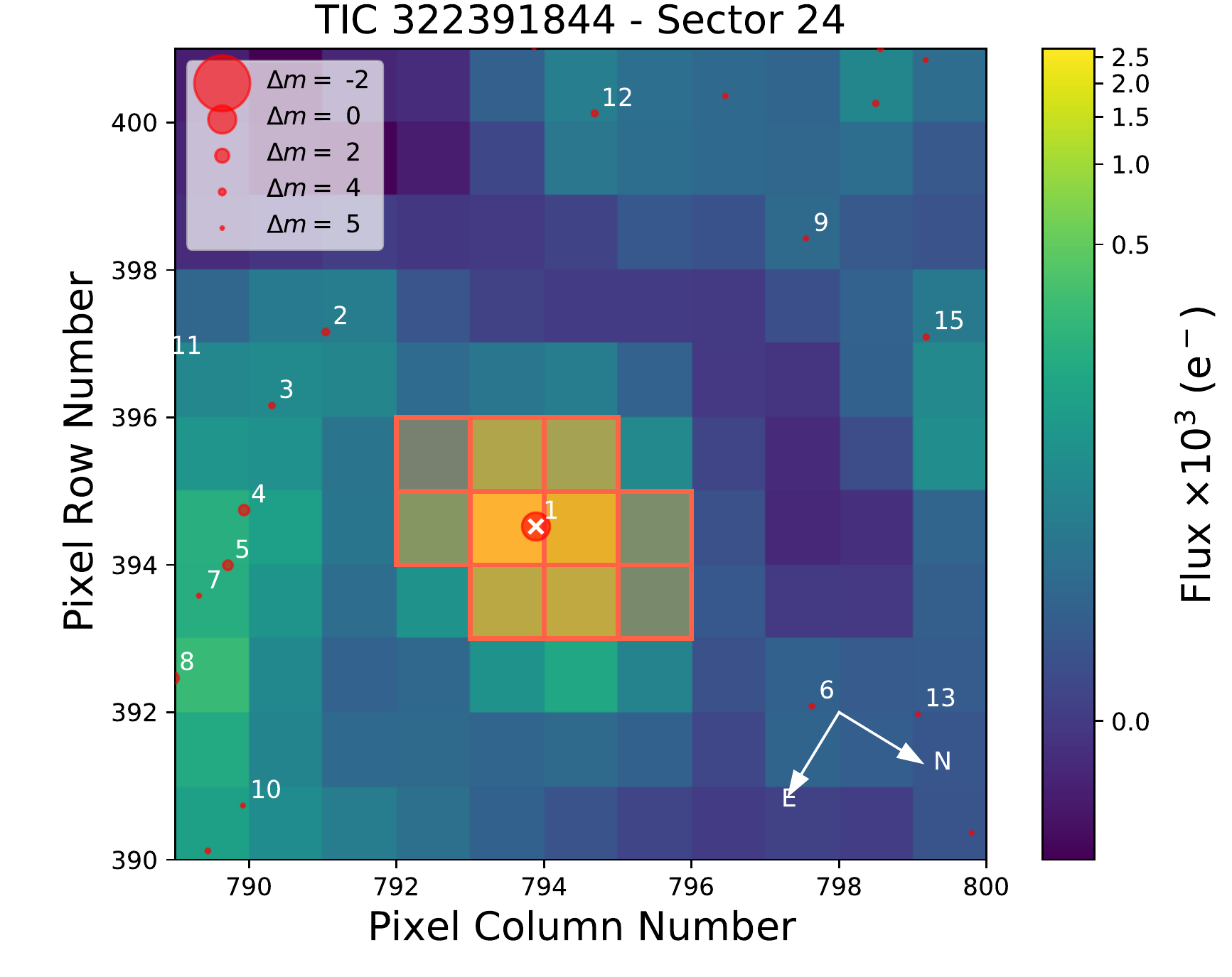} 
    \caption[]{TESS target pixel files of G\,264--012, marked with a cross at the centre of the plot, in sectors 16 (left), 18, and 22 (right). Red point represent {\em Gaia} DR2 sources down to 6\,mag fainter in $G$ than G\,264--012. 
    The orange regions enclose the different pixel masks used to extract the single aperture photometry. 
    {\em TESS} pixel size is about 21\,arcsec.}
    \label{fig:TPF_G264-012} 
\end{figure} 

\paragraph{SuperWASP.} 
The Super-Wide Angle Search for Planets\footnote{\url{https://www.superwasp.org}} \citep[SuperWASP,][]{Pollacco2006} consists of two robotic observatories, SuperWASP-North (La Palma, Spain) and SuperWASP-South (South Africa). Each has eight wide-angle cameras that simultaneously monitor the sky for planetary transits. The cameras, equipped with a 400\,nm to 700\,nm broadband filter, feed 2k$\times$2k CCDs with a FOV of 7.8$\times$7.8\,deg$^{2}$ per camera.
No data on G\,264--012 are available in the only SuperWASP public data release (DR1; Butters et al. 2010).
Data used here were acquired for Gl\,393 spanning the period from February 2008 to June 2014, including $\sim 74000$ useful observations distributed in clusters of $\sim 100$ days every season except in 2012. The data were reduced and detrended by the SuperWASP team following the algorithm described by \citet{Tamuz2005}. This algorithm corrects the light curve for trends that manifest in a collection of simultaneously observed stars, which are generally attributed to atmospheric extinction and detector systematics. 

\paragraph{MEarth.} 
The MEarth\footnote{\url{https://www.cfa.harvard.edu/MEarth}} project consists of two robotically controlled telescope arrays devoted to monitoring thousands of M-dwarf stars \citep{berta2012}. Since 2008, the MEarth-N (North) telescope array observes from the Fred Lawrence Whipple Observatory on Mount Hopkins (USA). This array consists of eight identical 0.4\,m robotic telescopes ($f$/9 Ritchey-Chr\'etien Cassegrain), each equipped with a 2k$\times$2k CCD camera, FOV of 26$\times$26\,arcmin$^2$, sensitive to red optical and near-infrared light. 
MEarth has generally used an RG715
long-pass filter, except for the 2010--2011 season when an $I_{715-895}$ interference filter was used. 
In the case of G\,264--012, three time series are available from the seventh data release, MEarth DR7, all collected with MEarth-N telescope T3.
We dubbed them MEarth-1 (October 2008 to May 2009), MEarth-2 (March to June 2011), and MEarth-3 (October 2011 to November 2015). 
 
\paragraph{ASAS-SN.} 
The All-Sky Automated Survey for Supernovae\footnote{\url{https://asas-sn.osu.edu}} (ASAS-SN) project \citep{Kochanek2017} currently consists of 24 telescopes distributed on six different sites around the globe. Each station consists of four 14\,cm aperture Nikon telephoto lenses, each with a thermo-electrically cooled, back-illuminated, 2k$\times$2k, Finger Lakes Instruments ProLine CCD camera with a wide FOV of 4.5\,deg$^2$.
Only one data set, for G 264-012, was publicly available.
This data set consisted in 601 useful V-band data points distributed over 222 nights during the period 2015-2018.

\paragraph{ASAS.} 
The All-Sky Automated Survey\footnote{\url{https://www.astrouw.edu.pl/asas}} project
\citep[ASAS,][]{Pojmanski1997}
has been monitoring the entire southern, and part of the northern, sky since 1997. It consists of two observing stations, at Las Campanas Observatory in Chile (ASAS-3 from 1997 to 2010, ASAS-4 since 2010) and Haleakal\={a} Observatory in Hawai'i (ASAS-3N since 2006). Both are equipped with two wide-field instruments observing simultaneously in the $V$ and $I$ bands. We retrieved 358 useful ASAS-3 $V$ measurements for Gl\,393 taken in the period between May 2001 and July 2009.


\paragraph{{\em Kepler}.} 
Gl\,393 was observed by the {\em Kepler Space Telescope} operating under the K2 mission\footnote{\url{https://keplerscience.arc.nasa.gov}} \citep{Howell2014} during campaign 14 from 2 June to 19 August 2017 with the long-cadence time resolution of $\sim$30\,min. 
The photometric light curves were retrieved from the Mikulski Archives for Space Telescopes\footnote{\url{https://mast.stsci.edu/portal/Mashup/Clients/Mast/Portal.html}} (MAST).
We used the pre-search data conditioning (PDC) light curves that are derived from processing the single aperture photometry (SAP) light curves by fitting co-trending basis vectors created from the stars in the same sector, camera, and CCD, to remove instrumental systematic variations and other perturbations such as pointing drifts, focus changes, gaps, or outliers. 
After removing outliers, mainly small flares, we reduced the number of 3537 useful data points to 3393.

\paragraph{{\em TESS}.}
G\,264--012 was observed with a 2\,min cadence by {\em TESS} in sectors S16 and S18 (cameras 2 and 3) in 2019 from 12 September to 6 October and 3 November to 26 November, respectively, and in sector S24 (camera 4) in 2020 from 15 April to 12 May\footnote{\url{https://heasarc.gsfc.nasa.gov/cgi-bin/tess/webtess/wtv.py?Entry=G+264-012}}. 
As in the case of Gl\,393 and {\em Kepler}, we used the PDC light curves retrieved from MAST, which are optimised to search for exoplanet transits.
The PDC light curves have 16\,812 (S16), 15\,334 (S18), and 18\,217 (S24) useful points each.
The three panels in Fig.~\ref{fig:TPF_G264-012}
show the target pixel files (TPFs) and SAP apertures used in sectors S16, S18, and S24.
We used \texttt{tfplotter}\footnote{\url{https://github.com/jlillo/tpfplotter}} \citep{Aller2020}, which overplots {\em Gaia} DR2 \citep{Gaia18a} sources in the nearby region. 
No objects down to 6\,mag fainter (less than 0.4\,\% in flux) in $G$ than G\,264--012 are visible in the {\em TESS} SAP photometric apertures.
{\em TESS} observed Gl\,393 in sector S35 during its extended mission between 9 February and 7 March 2021 with camera 1\footnote{\url{https://heasarc.gsfc.nasa.gov/cgi-bin/tess/webtess/wtv.py?Entry=Gl+393}}.

\section{Analysis and results}\label{sec:analysis}

\subsection{Signals in radial velocity data} \label{sec:signal_search}

To search for periodic signals, we computed the generalised Lomb-Scargle periodogram \citep{zech09} and determined the false alarm probabilities (FAP), as illustrated by Figs.~\ref{fig:RV_GLS_G264} and \ref{fig:RV_GLS_Gl393}. 
We considered a signal significant if its peak lies above the line for FAP = 0.1\,\% (i.e. with a FAP < 0.1\,\%). 
Significant signals were fitted and pre-whitened using Keplerian models. 
We continued searching in the pre-whitened RV data until no further significant signals were found. 

We used the insights from the periodogram analysis to apply more sophisticated models to the RV data.
Those models include planetary signals in the form of Keplerian orbits as well as the usage of Gaussian Processes \citep{Rasm06} to mitigate the influences of stellar noise.
These processes use a certain covariance function which is fitted to the autocorrelation function of the RV data.
In exoplanetary science, some of those so-called kernel functions consist of a periodic component, indicating the variation introduced by the rotating star, and an exponential decay which measures the stability over time of the signal \citep{celerite,Bau20,Tole21}.
We chose the commonly used quasi-periodic kernel \citep{2014MNRAS.443.2517H, 2015MNRAS.452.2269R, 2018MNRAS.474.2094A, 2020AJ....159...23N,2020ApJ...905..155G, Tole21} as shown here,

\begin{equation}
K_{\rm QP}(\tau) =  h^{2} \exp \Big( -\frac{\tau^{2}}{2 \lambda^{2}} - \frac{1}{2{\rm w}^{2}} \sin^{2}(\frac{\pi}{P} \tau) \Big), \label{Eq1}
\end{equation}

\noindent
which introduces the four hyper-parameters $h$, a parameterisation of the signals amplitude in m\,s$^{-1}$, a period $P_{\rm rot}$, describing the variability, the decay time of the correlation $\lambda$, and a factor $w$ relating those two timescales \citep{Per21}.
Together with the parameters of a Keplerian orbit (semi-amplitude $K$, orbital period $P$, time $t_{\rm c}$), offsets for the individual data sets $\mu$, and additional jitters $\sigma$, we minimise the likelihood function by exploring the parameter space with a Markov Chain Monte Carlo (MCMC), using the {\tt emcee} code \citep{Fore13} and the python-based {\tt george} code \citep{Amb15}.
A model is significantly preferred if the difference in log-likelihood exceeded $\Delta \ln{\mathcal L}$ = 15, in the sense that the higher the difference the higher the significance.
A difference of $\Delta \ln{\mathcal L} < 5$ is considered noise.
For a detailed discussion on the usage of this statistics for model comparison see \cite{Balu13}, and for an example
implementation of this procedure see \cite{Perger2019}.

\subsubsection{G\,264--012}

\begin{figure} 
    \includegraphics[width=\linewidth]{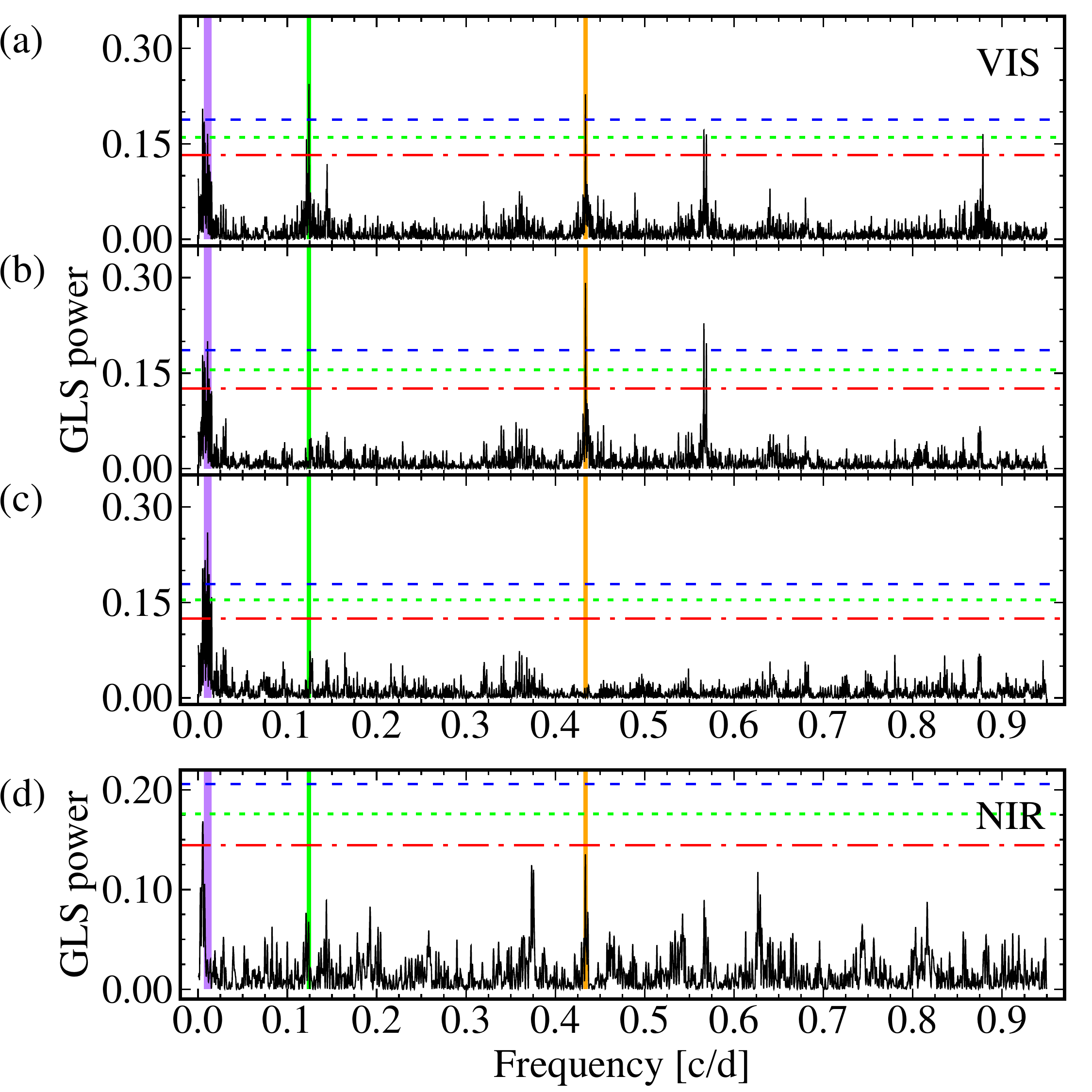} 
    \caption[]{GLS periodograms of G\,264--012 RVs.
    {\it Panel~($a$):} CARMENES VIS channel data. 
    {\it Panels~($b$) and ($c$):} residuals after pre-whitening the 8.05 and the 2.30\,d periods, respectively. 
    {\it Panel~($d$):} CARMENES NIR channel data.
    In the four panels, dashed vertical lines mark the signals at 92.70\,d (purple), 8.05\,d (green), and 2.30\,d (amber), while FAPs are indicated by horizontal lines: 10\,\% (red dash-dotted), 1\,\% (green dotted), 0.1\,\% (blue dashed).}
    \label{fig:RV_GLS_G264} 
\end{figure}

\begin{table}
	\centering 
	\caption{{Log-likelihood $\ln{\mathcal L}$ for different RV models for G\,264--012 and Gl\,393. Black font indicates the preferred solution}} 
	\label{tab:RV_signals} 
	\begin{tabular}{l cc cc}
		\hline \hline \noalign{\smallskip}
	$N_{\rm planets}$ &  \multicolumn{2}{c}{G\,264--012} & \multicolumn{2}{c}{Gl\,393}\\	
	&  no GP & GP &  no GP & GP\\
		\noalign{\smallskip} 
		\hline 
		\noalign{\smallskip}
	0 & --456.4 & --424.1 & --807.6 & --727.4\\ 
	1 & --443.0 & --422.6 & --799.6 & --706.8\\
	2 & --422.6 & --393.7 & --798.6 & --705.4\\
	3 & --413.4 & --389.5 & ... & ... \\
		\noalign{\smallskip}
		\hline 
	\end{tabular} 
\end{table}

Using our approach, we found that the G\,264--012 CARMENES VIS RVs support the presence of three signals ($P_{b}=2.30$\,d, $P_{c}=8.05$\,d and $P_{d}=92.70$\,d; see panels ($a$--$c$) in Fig.~\ref{fig:RV_GLS_G264}). 
The two highest peaks are also found in the CARMENES NIR data, at $215.6$\,d (probably related with twice $P_{d}$ in the VIS channel, considering the frequency resolution of 0.00076\,c\,d$^{-1}$ corresponding to our RV data set, given by the time span of $\Delta T=1316$\,d, see Sect. 3.1) and $2.30$\,d (same as $P_{b}$), though none of them were significant, as shown in panel ($d$) of Fig.~\ref{fig:RV_GLS_G264}.

A comparison of $\ln{\mathcal L}$ can be found in Table~\ref{tab:RV_signals}.
The values show that a 2-planet + GP model is the preferred one. 
They also indicate that modelling the long period signal $P_{d}$ with a Gaussian process instead of a Keplerian fit was highly preferred. 
This hinted towards stellar activity as the origin for the $92.70$\,d signal, but we gathered more evidence for this hypothesis in Sects.~\ref{sec:coherence}, \ref{sec:rotation}, and \ref{sec:activity_indicators}, where we analyse photometric data and spectroscopic activity indicators in G\,264--012. These indicators show indeed that this long-period signal is most probably due to activity. In Sect.~\ref{sec:combined_analysis}, we study the system with a combined analysis with Gaussian Processes.

\subsubsection{Gl\,393}

\begin{figure}
     \includegraphics[width=8.5cm]{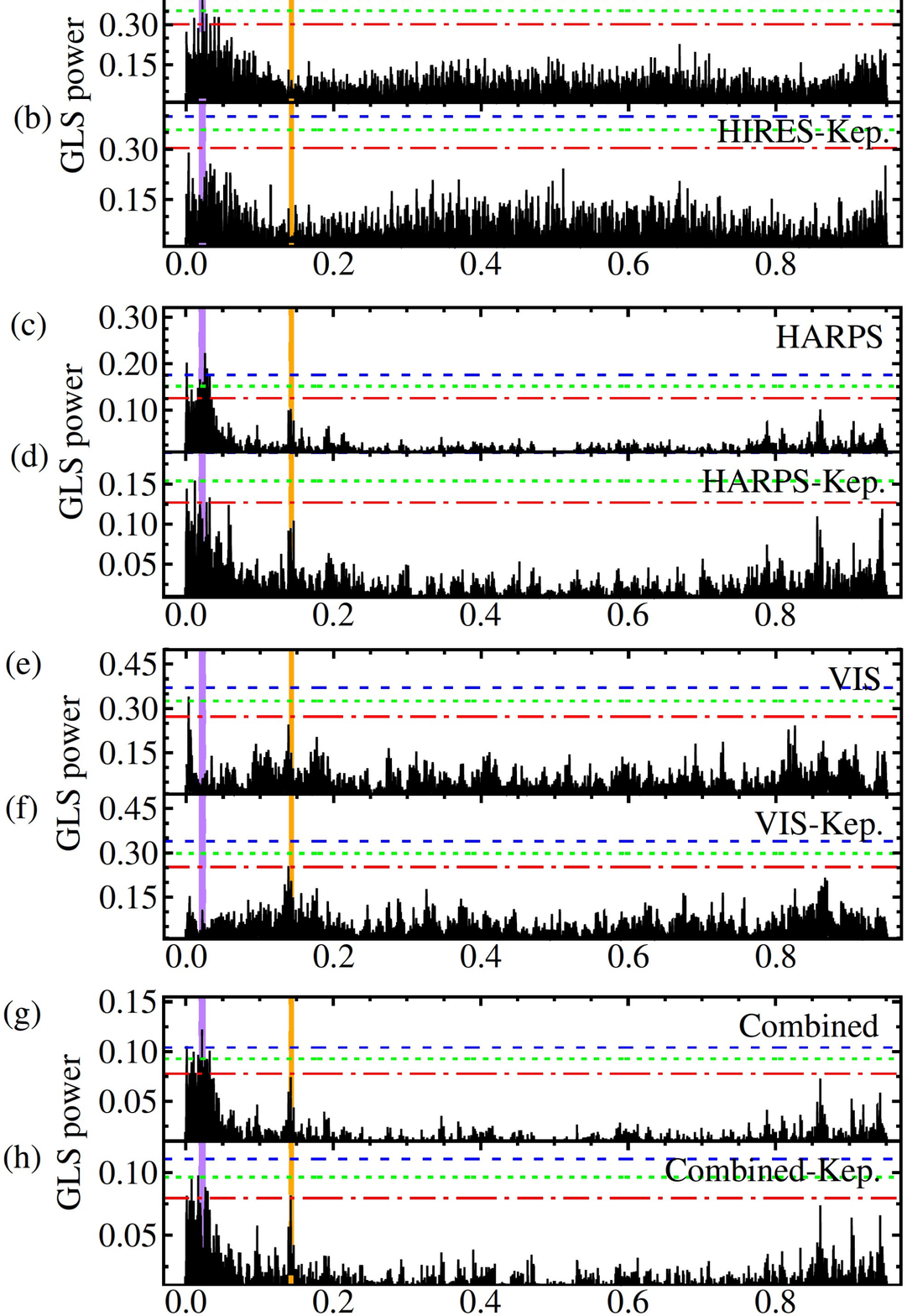}
     \caption[]{GLS periodograms for Gl\,393: {\it Panel~($a$):} HIRES data, and {\it Panel~($b$):} HIRES data residuals after pre-whitening the highest peak with a Keplerian fit. {\it Panels~($c$), ($d$):} HARPS data (panel descriptions as in those for HIRES), {\it Panels~($e$), ($f$):} CARMENES VIS channel data, and {\it Panels~($g$), ($h$):} combined data set. 
     FAPs as in Fig.~\ref{fig:RV_GLS_G264}.
     Vertical dashed lines mark the highest peak found in the combined data (purple) and the highest peak after pre-whitening the GP (orange).}
     \label{fig:RV_GLS_Gl393} 
\end{figure}

The GLS periodograms for individual instruments, as well as the combined RV data set, of Gl\,393 can be seen in Fig.~\ref{fig:RV_GLS_Gl393}. 
In the HIRES (panel $a$) and HARPS (panel $c$) data sets, we found significant periods at 45.6 and 38.5\,d, respectively, which appear within a forest of low frequency peaks. 
In the VIS channel data of CARMENES, no significant signal was found (panel $e$). 
The highest peak represents a period of $302$\,d and reaches a FAP of 0.5\,\%.
When combining all data sets (panel $g$), we found a significant period at 45.5\,d, again within a forest of low-frequency peaks. 
After fitting and subtracting a Keplerian signal to the respective frequencies, we did not find a formally significant peak in any individual (panels $b$, $d$, and $f$) or in the combined (panel $h$) data set.
However, we noticed the presence of a relatively high FAP ($\approx 10$\,\%) signal at 7.2\,d in the VIS channel and at 7.0\,d in the HARPS data and in the combined data.
The signal is not seen in the HIRES data probably because of its higher rms scatter.

To investigate the stability of this second isolated frequency, we split the HARPS data set into two chunks of 90 data points each and repeated the periodogram analysis.
We found the signal at around 7.0\,d to be present in both HARPS subsets, while the forest of peaks around 38\,d was highly variable and disappeared in the second HARPS subset. 
We also found the forest in the CARMENES data, but with a FAP barely below 10\,\%.
Thus, we attributed the power seen at low frequencies in the periodograms to stellar activity, which is variable over time.

The complex nature of stellar activity hinders any signal search based on classical Keplerian pre-whitening approaches. 
Instead, we fit a Gaussian process (GP) with a quasi-periodic kernel to the combined RV data set of CARMENES, HARPS, and HIRES.
We tested several models against the Null model (no signals): a number of Keplerian models (around 7.0\,d), a GP, and combined Kepler+GP.

As shown in Table~\ref{tab:RV_signals}, the GP is highly preferred over a Keplerian fit.
Its significance also continuously increases with time when the RV points are sequentially added to the fit, showing that the RV data is dominated by complex magnetic activity.
However, as we added additional RV data, the GP activity model alone is not enough to fit the data.
When the combined RV data was corrected with the GP model (see panel $g$ of Fig.~\ref{fig:RV_GLS_Gl393}), a significant (FAP = 0.1\,\%) signal was found at 7.0\,d.

The $\ln{\mathcal L}$ comparison resulted in a highly preferred fit when a one-planet Keplerian model was combined with the GP, with subsequent addition of planets resulting in negligible improvement (see also Table~\ref{tab:RV_signals}). 
Hence, the signal found around 7.0\,d was ascribed to be of planetary origin.
Nevertheless, we investigated the photometric data and spectral activity indicators and provide our results in Sects.~\ref{sec:rotation} and~\ref{sec:activity_indicators}, respectively, where we show that the origin of the other signals is most probably due to activity.

\subsection{Temporal and chromatic coherence}\label{sec:coherence}

Testing the temporal and chromatic coherence of signals is a powerful tool to distinguish planetary Doppler signals from stellar activity with RV data alone \citep{Bau20}. 
In practice, these tests are often difficult to conduct because they require a large amount of RV data, especially in the case of low amplitude signals.

\subsubsection{G\,264--012}

\begin{figure} 
     \includegraphics[width=\linewidth]{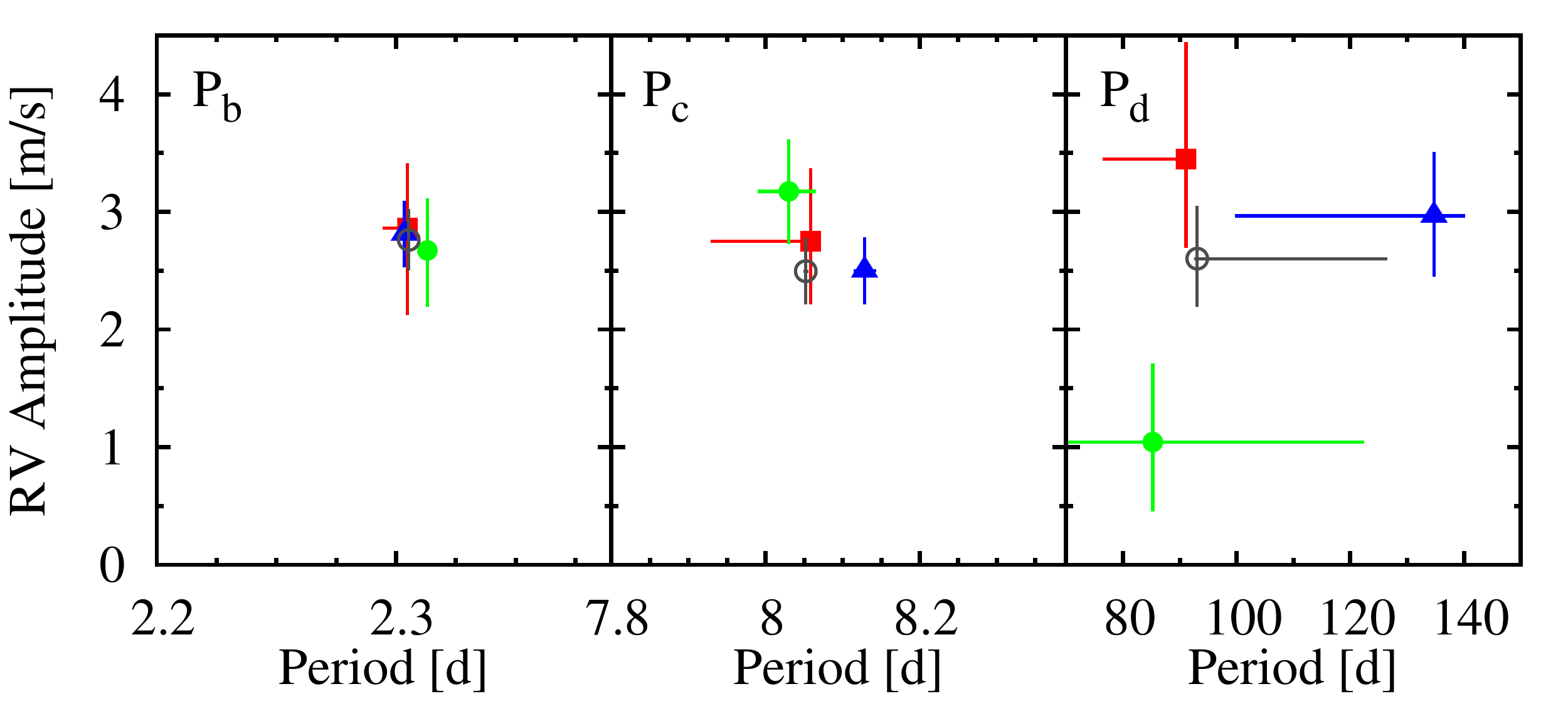}
     \includegraphics[width=\linewidth]{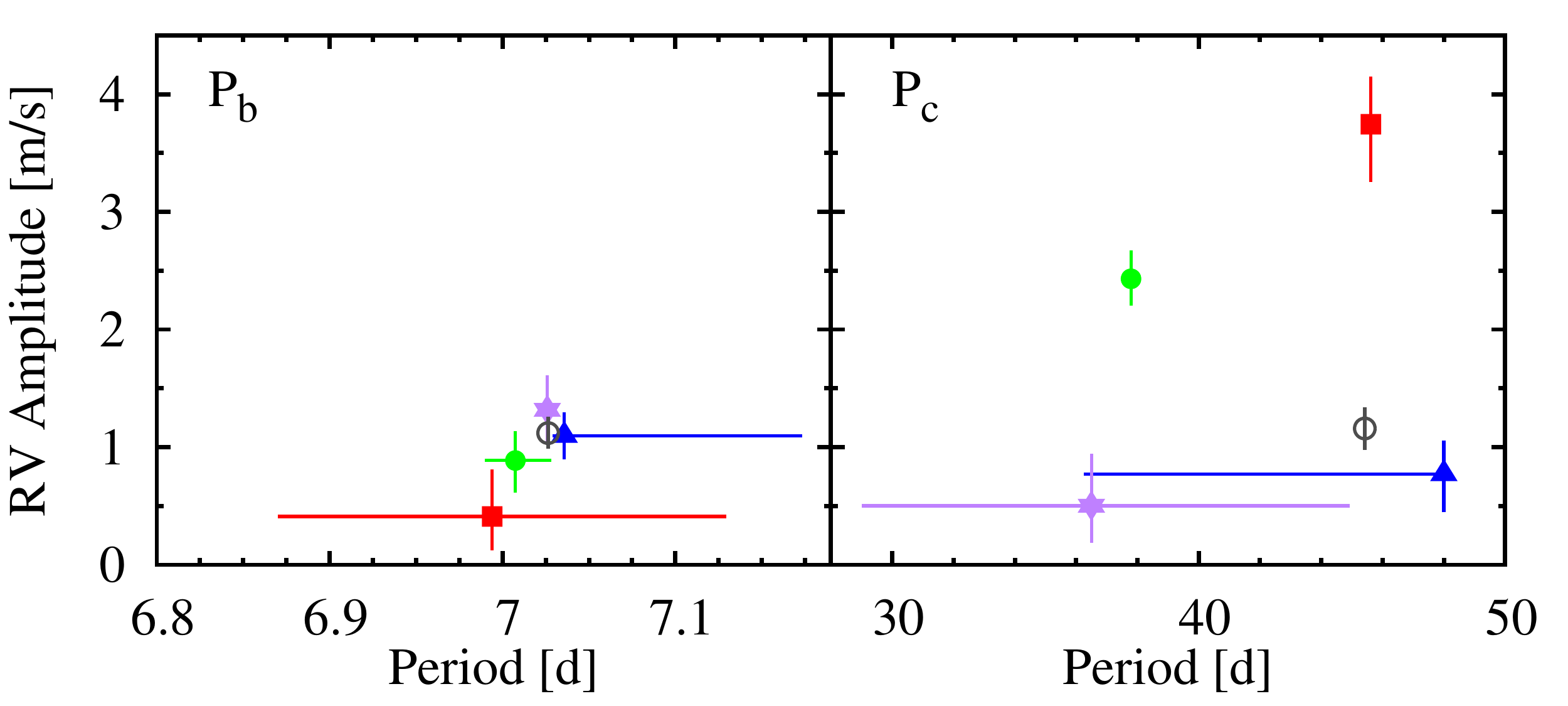}
     \caption[]{{\em Top panels}. Period-amplitude diagrams of the Keplerian fits for the three signals in the G\,264--012 RV data set split into seasons: 2016--17 (red squares), 2018 (green points), 2019 (blue triangles), and all data (grey circle). 
     {\em Bottom panels}. Same as the top panels but for the Gl\,393 RV data split into subsets: HIRES (2008--2013; red squares), HARPS subset 1 (2003--2014; green points), HARPS subset 2 (2014--2017; blue triangles), CARMENES VIS (2016--2019; purple star), all data (grey circle).}
     \label{fig:phase_cohe} 
\end{figure}

We started by analysing the temporal coherence of all three signals found in the VIS channel data of G\,264--012. 
We split the VIS data set into seasons, where RVs taken during 2016 and 2017 were treated as one season because the sampling was too sparse to reliably fit three Keplerian orbits to each single season. 
The data taken during 2018 and 2019, however, were dense enough to treat them independently, as shown in Fig.~\ref{fig:phase_cohe}, top panel. 
The two signals found at $P_{b}=2.30$\,d and $P_{c}=8.05$\,d were relatively stable in amplitude and period between different observing seasons, strengthening the planetary origin hypothesis. 
However, the signal at $P_{d}=92.70$\,d was variable in both amplitude and period over different seasons. Especially during the 2018 season the signal almost vanished in the RV data, which is a strong indication for variability due to stellar activity \cite[cf.][]{Bau20}.

Chromatic coherence was difficult to test in G\,264--012 because we lacked sufficient precision for the low amplitude signals to be seen with CARMENES NIR. 
Nonetheless, at least for the shortest period signal at $P_{b}=2.30$\,d, we could carry out a chromatic test of indicative character. 
We fixed the period to the one found with the VIS channel and fit both VIS and NIR RV data sets independently with a circular Keplerian orbit. 
Both fits yielded consistent RV semi-amplitude results, $K$, both in CARMENES VIS and NIR ($K_{\mathrm{VIS}} = 2.76 \pm 0.26$\,m\,s$^{-1}$, $K_{\mathrm{NIR}} = 1.88 \pm 1.20$\,m\,s$^{-1}$), as expected from a planetary signal origin. 

\subsubsection{Gl\,393}

Testing chromatic and temporal coherence in the RV data of Gl\,393 was also challenging, especially for the signal at 7.0\,d, as its amplitude is only about 1\,m\,s$^{-1}$. 
To test the temporal coherence, we analysed the HIRES, HARPS, and CARMENES VIS data sets independently, but split the large HARPS set into two subsets of 90 data points each. 
In each data set, we fit both, a signal between 25 and 50 days and one at around 7.0\,d. 

The signal around 45.5\,d (right panel in Fig.~\ref{fig:phase_cohe}) decreased in amplitude over time.
It exhibited an amplitude of about 4\,m\,s$^{-1}$ in the HIRES data and decreased to 2.5\,m\,s$^{-1}$ in the first HARPS subset before later vanishing completely in the second HARPS subset and the CARMENES data (in the last two cases the derived amplitude fell below the instrumental precision at around 1\,m\,s$^{-1}$). 
The behaviour observed in this signal points towards variability due to stellar activity.
The amplitudes and periods around 7.0 days derived from CARMENES and HARPS data are quite consistent, whereas the HIRES data yield a non-detection (possibly due to the larger rms and fewer RV data points). 
Nonetheless, the derived RV amplitudes of about of 1\,m\,s$^{-1}$ are around the limit of a reliable detection from all subsets (see left panel of Fig.~\ref{fig:phase_cohe}).
We can, therefore, extract no information regarding the temporal coherence of the 7.0\,d RV signal from this test.

Testing chromatic coherence in Gl\,393 is beyond what can be achieved with the current RV data set.
This is the case because CARMENES did not observe Gl\,393 long enough to reach the required precision with its NIR channel. 

\subsection{Search for variability in photometric data}\label{sec:rotation}

\begin{figure*}
    \centering
    \includegraphics[width=\textwidth, trim = 2cm 6.9cm 2cm 2cm, clip]{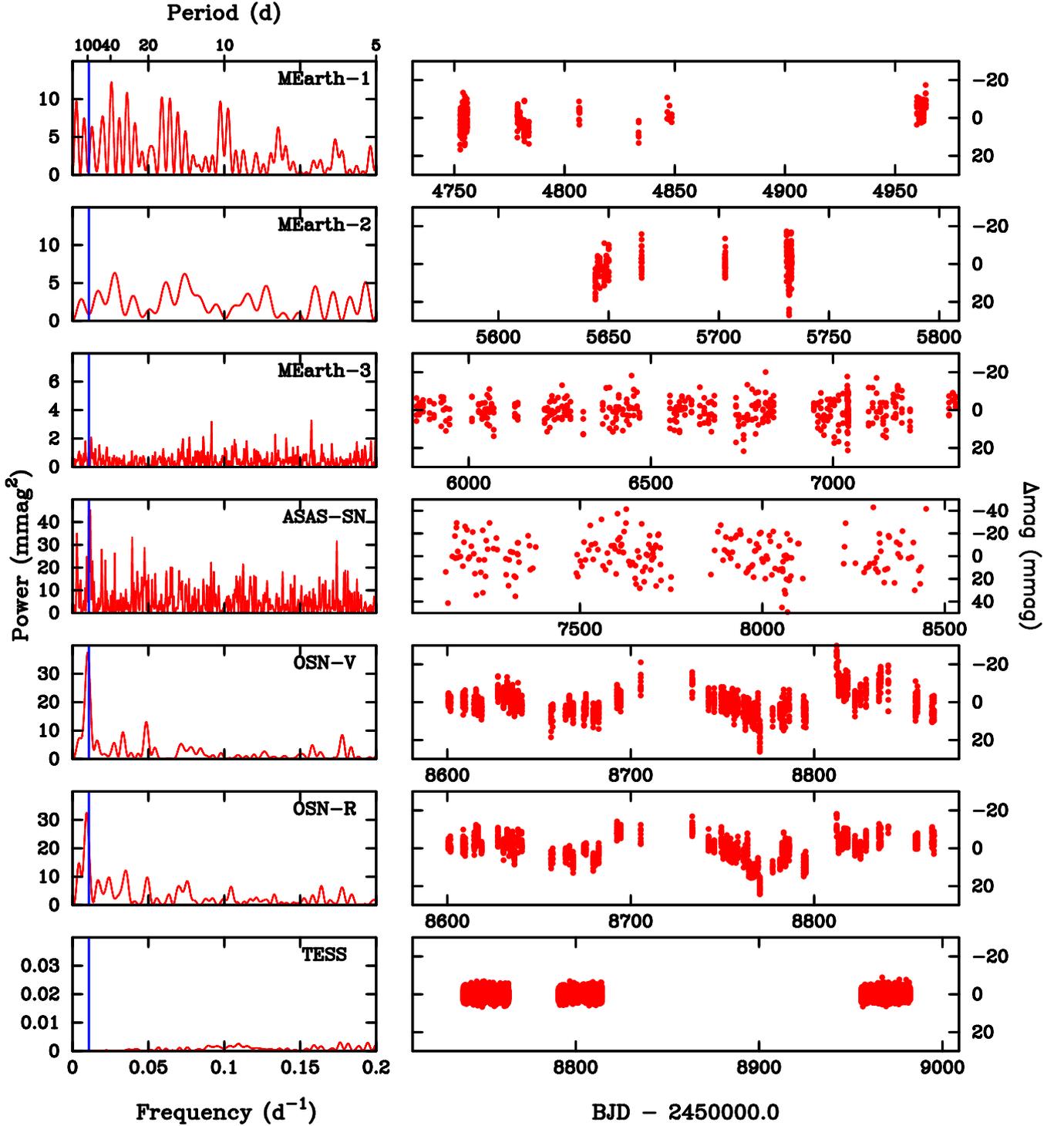}
     \caption[]{Analysis for periodicity using available photometric data sets of G\,264--012. 
     {\em Left panels}: power spectra.
     {\em Right panels}: time series. 
     The vertical blue solid line in the left panels indicate the period of 92.7\,d derived in the RV time series.}
     \label{fig:photometry_frequency_analysis} 
\end{figure*} 

In single M dwarfs, brightness variations are caused by starspots, or other activity structure on the stellar surface, moving across the stellar disc by rotation.
The two host stars, G~264-012 and Gl~393, were monitored in photometric surveys in search for these variations to determine their rotation periods.


\citet{Newton2016} analysed the first 230 MEarth data points of G\,264--012 collected in the 2010 campaign.
They determined $P_{\rm rot}$ = 48.51\,d with $A$ = $9.9\pm4.1$\,mmag but considered it a `non-detection or undetermined detection' (source type N) with a `bright contaminant' (contamination flag 1).
\citet{DiezAlonso2019} extended the analysis to 790 MEarth data points collected in the campaigns 2010, 2011, and 2014, but did not find a reliable period, either.

The ASAS light curve of Gl\,393 was also investigated by \citet{DiezAlonso2019}.
In spite of the 413 data points over 8.636\,a, they again did not find any reliable period.
Later, \citet{Rein20} determined $P_{\rm rot}$ = 35$\pm$13\,d with a variability range of 0.49\,\% using ultra-precise {\em Kepler} data collected during the K2 campaign 14 (between 31 May and 19 August 2017).

The undetermined period of G\,264--012 and the large uncertainty in the Gl\,393 rotation period led us to perform our own search for low-amplitude photometric variability, including a brand-new transit search.
For the two stars, we carried the transit search with the transit least squares ({\tt TLS}) algorithm \citep{TLS}, which is optimised for the detection of small planets with low signal to noise in large data sets, such as those supplied by {\em TESS} or {\em Kepler}. 
{\tt TLS} evaluates a statistic, the signal detection efficiency (SDE), to identify transiting candidates. 
Typically, an SDE = 8 is equivalent to an FAP $\approx$ 10\,\%.

\subsubsection{G\,264--012}

\begin{figure} 
     \includegraphics[width=\linewidth]{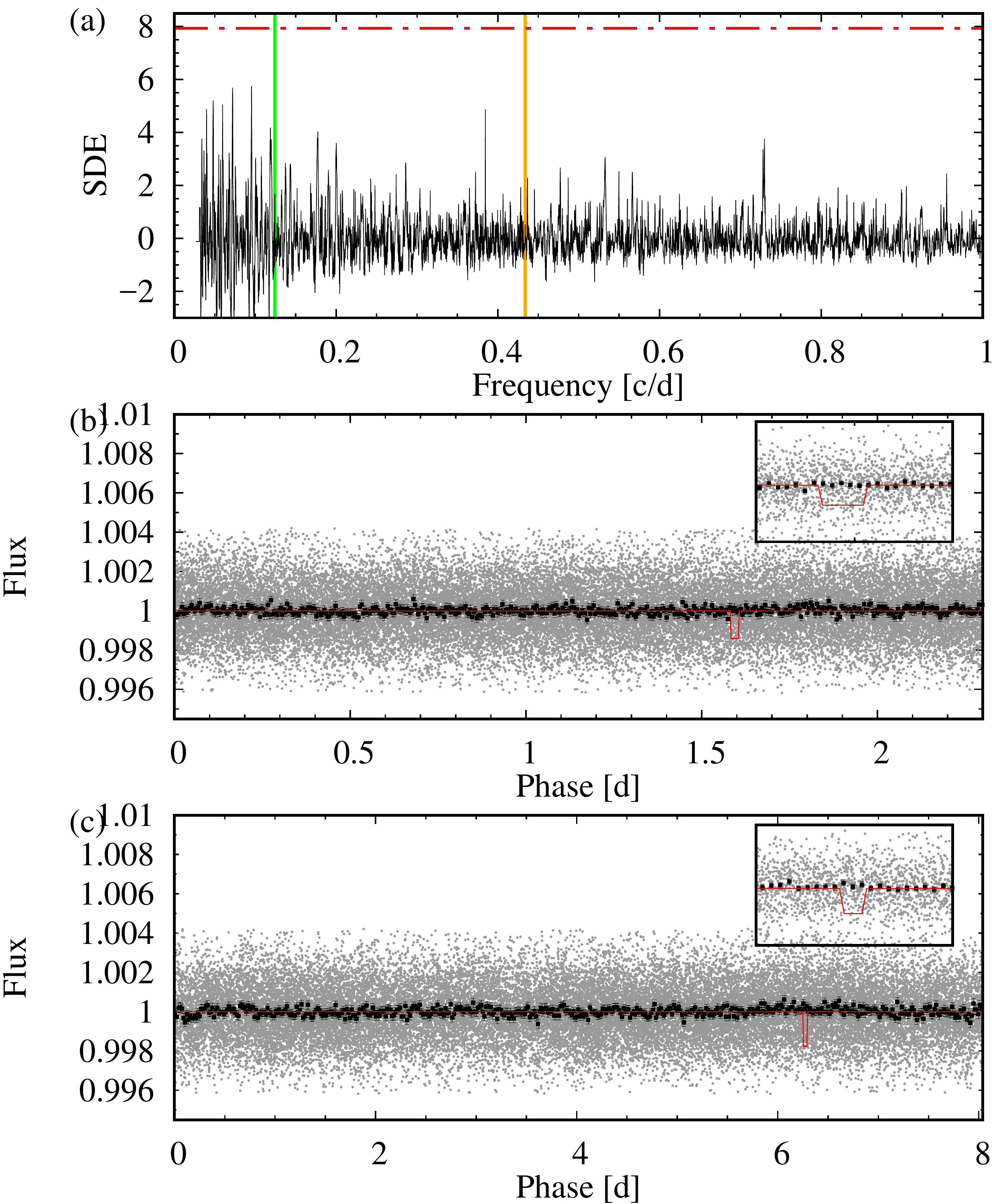} 
     \caption[]{{\it G\,264--012: Panel~($a$):} transit search signal detection efficiency (SDE) as a function of frequency using the {\tt TLS} algorithm. The red dashed-dotted horizontal line marks the 10\,\% FAP, the vertical green and yellow lines mark the signals found in the CARMENES RV data at 8.05\,d and 2.30\,d, respectively. 
     {\it Panel~($b$):} \emph{TESS} data folded to the $2.30$\,d signal. \emph{TESS} two minute cadence data (grey dots), phase binned \emph{TESS} data (black squares), expected transit signal (solid, red line). 
     {\it Panel~($c$):} same as panel ($b$) but for the $8.05$\,d signal.}
     \label{fig:TESS} 
\end{figure} 

We analysed all the available photometric data sets of G\,264--012, including the new and contemporaneous measurements collected at OSN.
The results are summarised in Fig.~\ref{fig:photometry_frequency_analysis} and Table~\ref{table:photometry}. 

As \citet{Newton2016} and \citet{DiezAlonso2019}, we did not find any significant periodicity in the frequency analysis of the MEarth data, whose periodogram shows a very low mean power level, suggesting a very low activity level during the epoch of these observations (2010--2015).
However, the level of surface activity in G\,264--012 increased after 2015.
Although the ASAS-SN data set is noisier than MEarth ones, its periodogram shows power in the low-frequency region, with a main peak at frequency $f$ = 0.01078\,d$^{-1}$ ($P$ = 92.8\,d, amplitude $A \approx$ 7.0\,mmag), consistent with the long-period signal of 92.7\,d detected in the RV time series.
The same results were obtained after nightly binning the ASAS-SN time series.
Furthermore, the analyses of the OSN photometric time series in both $V$ ($P \sim$ 99.7\,d) and $R$ ($P \sim$ 107.5\,d) filters were consistent with each other and with the ASAS-SN data set.

The {\em TESS} data set consists of a time series obtained during three sectors, S16, S18 and S24, with effective time spans of 24.7, 23.7 and 26.5\,d, respectively, and a total time baseline of 244\,d.
In general, {\em TESS} SAP light curves are commonly affected by instrumental drifts, while the PDC ones are not.
However, when creating the PDC light curves, low-amplitude, mid-term variations intrinsic to the star are also disturbed or even removed.
In the case of G\,264--012, the relatively long periodicity of about 90--100\,d, the small amplitude of a few millimagnitudes, and the time gaps of nearly 60\,\% of the baseline prevented us from detecting any rotational modulation in the {\em TESS} data.
In particular, we imposed an upper limit of 0.15\,mmag to the amplitude of the stellar variability remaining in the PDC light curve (Fig.~\ref{fig:photometry_frequency_analysis}), as given by the maximum level of the forest of peaks in the noise shown in its periodogram.

We used the {\tt george} kernel \citep{Amb15} and a dynamic nested sampling with {\tt juliet} \citep{Espinoza2019} for a GP fit on the combined, nightly-binned, photometric data. 
Of the four GP hyperparameters of this kernel, the amplitude and harmonic complexity were individual for each data set, whereas the rotation period and decay timescale were shared by all of them.
 
After injecting a uniform $P_{\rm rot}$ prior between 1 and 200\,d, we measured a posterior rotation period of 112.5$^{+1.2}_{-1.5}$\,d. 
We repeated the fit after discarding the MEarth and {\em TESS} data sets and keeping only the ASAS-SN and OSN sets, with the clearest signals, and found a consistent value with a smaller uncertainty, $P_{\rm rot;GP}$ = 112.6$\pm$0.9\,d.

Given the low amplitude of the photometric variability of G\,264--012, of the order of the error bars of the individual data points, we were conservative in assigning a mean rotation period for the star of 100$\pm$6\,d, as derived from the three values presented in Table~\ref{table:photometry}.

We looked for transits in the {\em TESS} sectors.
As illustrated by Fig.~\ref{fig:TESS}, no transit is evident from the SDE at the periods of 2.30\,d and 8.05\,d found in CARMENES RVs (Sect.~\ref{sec:signal_search}).
The position of the transits were computed from the RV solution, while the transit depths were estimated assuming a rocky and silicate composition \citep{Zeng_2019}, as corresponding to hypothetical Earth-like planets.
The expected transit depths were 1500\,ppm for the 2.30\,d signal and 1900\,ppm for the 8.05\,d signal.
As a comparison, the {\em TESS} light curve has an rms of 654\,ppm and error bars of typically 766\,ppm.

\subsubsection{Gl\,393}

\begin{figure*}
    \centering
    \includegraphics[width=\textwidth, trim = 2cm 11.5cm 2cm 2cm, clip]{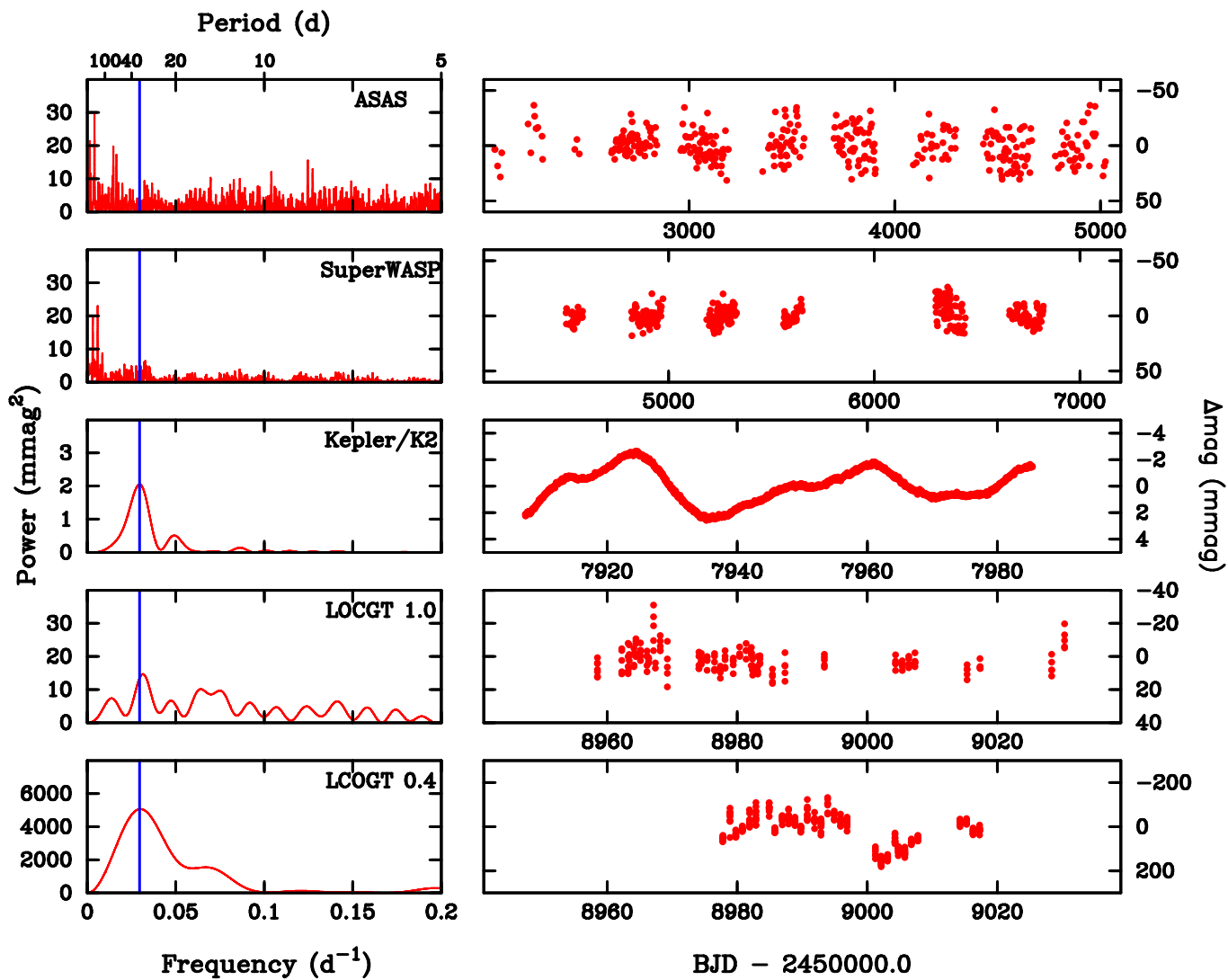}
     \caption[]{Same as Fig.~\ref{fig:photometry_frequency_analysis}, but for Gl\,393 and the period of 34.0\,d derived from the {\em Kepler} K2 light curve.}
     \label{fig:photometry_frequency_analysis_GJ393} 
\end{figure*} 

\begin{figure} 
     \includegraphics[width=\linewidth]{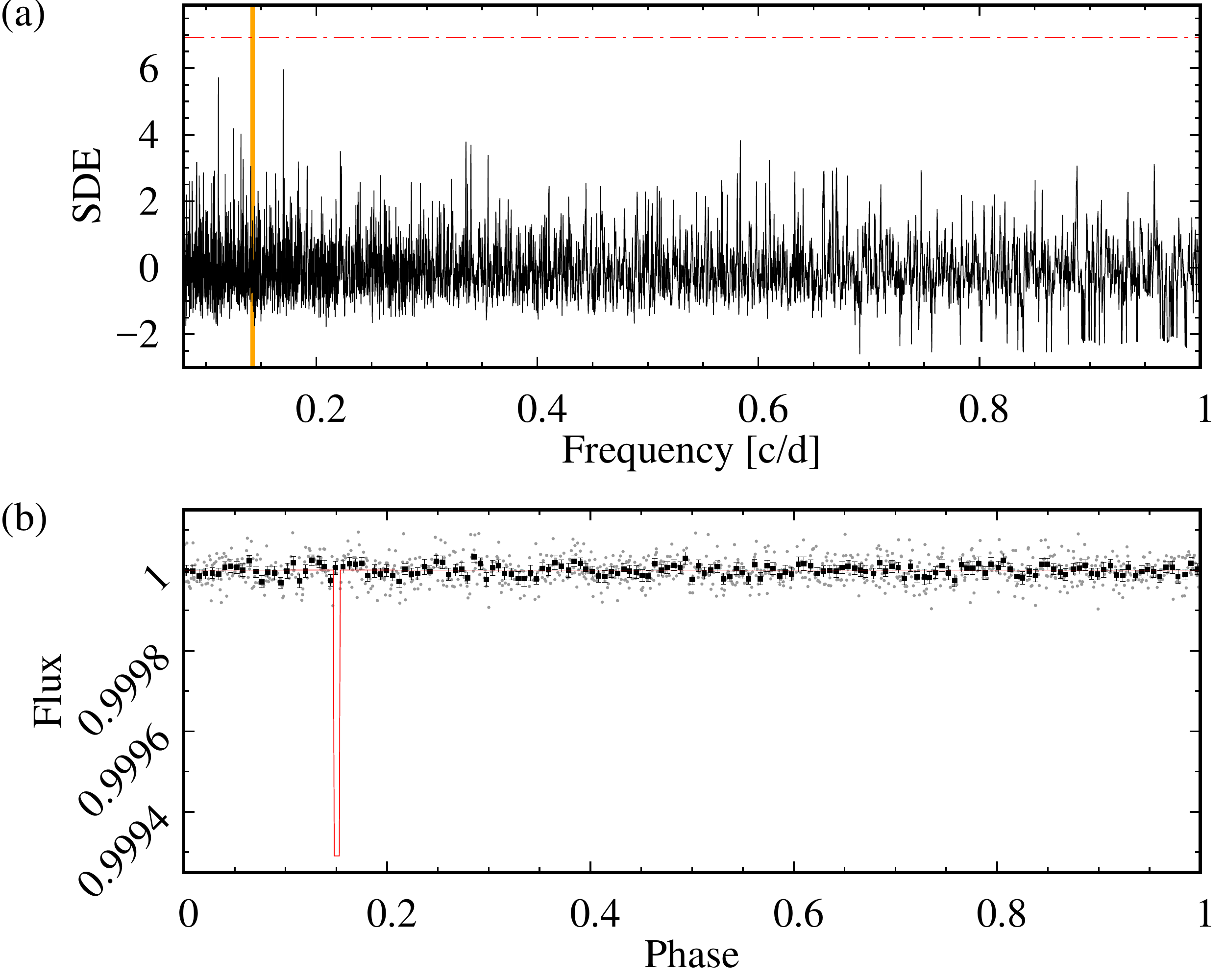}
     \caption[]{Gl\,393: {\it Panel~($a$):} Transit search signal detection efficiency (SDE) as a function of frequency for K2 data. The horizontal dashed-dotted line marks the 10\% FAP, while the vertical yellow dashed line marks the 7.0\,d signal found in the RVs. {\it Panel~($b$)}: K2 data (grey) folded to the 7.0\,d signal. Black dots are the binned K2 data, while the red line shows the simulated transit.}
     \label{fig:K2_transit_fit} 
\end{figure}

The high-quality {\em Kepler} K2 light curve (see Sect.~\ref{sec:combined.Gl393}) was analysed in order to determine the rotation period of Gl\,393.
The light curve displays flares, as expected from a moderately active star, on top of significant quasi-periodic variations with time scales of tens of days and a long-term linear trend, probably of instrumental origin.
Hence, we first linearly detrended the light curve and removed flares with a $3 \sigma$-clipping filter.
Our frequency analysis, with the interactive computer package {\tt Period04} \citep{LenzBreger2005}, which is based on algorithms similar to the Lomb-Scargle periodograms, resulted in a main periodicity of $P \sim 34.0\pm0.1$\,d.
The amplitude of this variability is $A = 3.0\pm0.1$\,mmag (full peak-to-peak amplitude), which we attributed to the rotation period of Gl\,393.
This is in good agreement with the determination of $35.1\pm13.0$\,d by \citet{Rein20}.
Furthermore, there is no sign of any peak at 7.0\,d that could challenge the planetary origin of this signal.

As expected from the very small amplitude of the surface activity signal shown by the {\em Kepler} K2 light curve of Gl\,393, we did not find any significant periodicity in the frequency analysis of the time series obtained from ground-based photometry alone (ASAS, SuperWASP, and LCOGT; see Fig.~\ref{fig:photometry_frequency_analysis_GJ393} and Table~\ref{table:photometry} for a summary).
However, the combination of different data sets may improve the $P_{\rm rot}$ determination.
Thus, we made joint fits with a quasi-periodic GP with different data set combinations.
Probably due to the relatively large uncertainty of the ASAS and SuperWASP data with respect to the small amplitude of variability, the combination that led to the most robust $P_{\rm rot}$ determination was that of joining {\em Kepler}, LCOGT 1.0\,m, and LCOGT 0.4\,m, which resulted on a rotation period $P_{\rm rot}$ = 34.15$^{+0.22}_{-0.21}$\,d (68\,\% confidence interval uncertainties), similar to that of \citet{Rein20}.

We performed a transit search in the available K2 data of Gl 393 (Fig.~\ref{fig:K2_transit_fit}). 
For that, after detrending and flare-subtraction, we modelled and corrected the light curve with a GP with M\'atern kernel prior to its analysis with {\tt TLS}. 
The 7.0\,d signal would produce a transit depth of 770\,ppm assuming, as in G264-012, a terrestrial composition for the planet.
Such a transit signal should be clearly visible in the K2 light curve, which has an rms of 35\,ppm and error bars of typically 38\,ppm.
However, we do not find any evidence for a periodic dip. 

\subsection{Search for variability in the activity indicators}\label{sec:activity_indicators}

\subsubsection{G\,264--012}

\begin{figure} 
     \includegraphics[width=\linewidth]{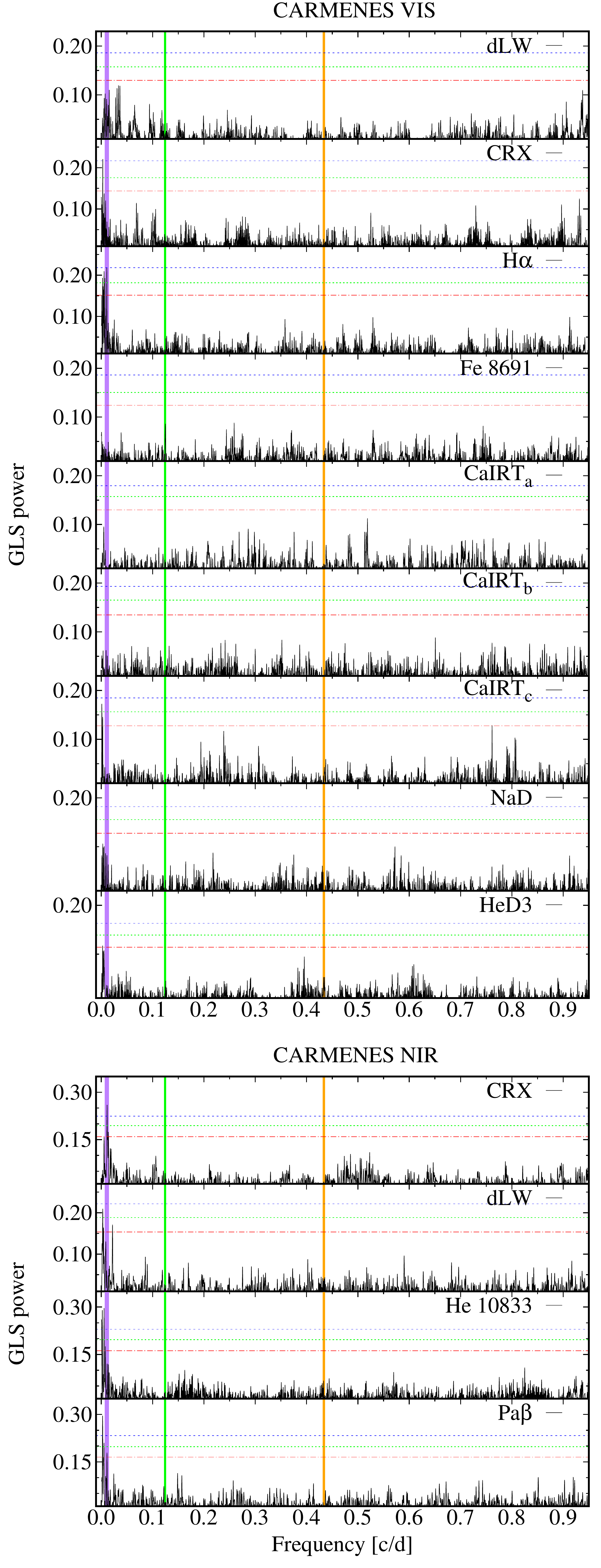} 
     \caption[]{GLS periodograms of CARMENES activity indicators of G\,264--012 in the VIS ({\em top panels}) and NIR ({\em bottom panels}) channels. 
     Horizontal FAP lines and vertical signal indications are the same as in Fig.~\ref{fig:RV_GLS_G264}.}
     \label{fig:GLS_activity_G264} 
\end{figure} 

In addition to the photometric follow up, we also investigated a number of spectral line indices that are sensitive to photospheric or chromospheric activity.
All of them were already successfully used, or recently suggested in the literature \citep{Na_in_M_stars,CaIRT_as_activity_indicator,Scho19,rotation_in_16_mdwarfs}, to uncover stellar activity signals.
We also used the {\serval} indicators, differential line width (dLW) and chromatic index (CRX) defined by \citet{Zech18} to gain further insights into the nature of the signals found in the RVs and the photometry.
In Fig~\ref{fig:GLS_activity_G264}, we present the GLS periodograms of all our activity indicators gathered for G\,264--012.

Among the collected spectroscopic indicators, we found H$\alpha$ ($P=92.4$\,d), He~{\sc i} $\lambda$10\,833\,{\AA} ($P=183.3$\,d), Pa$\beta$ ($P=215.7$\,d), VIS CRX ($P=339.4$\,d), and NIR CRX ($P=88.2$\,d) to exhibit significant signals above the $\mathrm{FAP}=0.1$\,\% level.
The periods found in H$\alpha$ and NIR CRX match the signal $P_{d}=92.7$\,d found in the VIS channel RVs. 
Furthermore, He~{\sc i} $\lambda$10\,833\,{\AA} and Pa$\beta$ reflect the highest peak found in the NIR channel RVs ($215.6$\,d) and are, thus, likely associated with $P_{d}$ via a factor of $2$, taking into account the frequency resolution of 0.00076\,c\,d$^{-1}$ corresponding to our RV data set, from a time span of $\Delta T=1316$\,d (see Sect. 3.1).
An even better agreement can be found if we assume that the true $P_{\rm rot} \sim 100$\,d for G-264-012, as follows from the photometric results listed in Table 3.
That is, the frequency resolution is not enough to resolve between the two resulting values (215\,d and 2$P_{\rm rot}$) in our periodograms
The signal of 339.4 d found in the VIS CRX periodogram is close to a one-year periodicity and therefore may represent an alias.
Using all the evidence collected from photometry (i.e. ASAS-SN and OSN) and spectroscopic activity indicators, we concluded that the RV signal $P_{d}=92.7$\,d is linked to the rotational period of G\,264--012.

\subsubsection{Gl\,393}

\begin{figure} 
     \includegraphics[width=\linewidth]{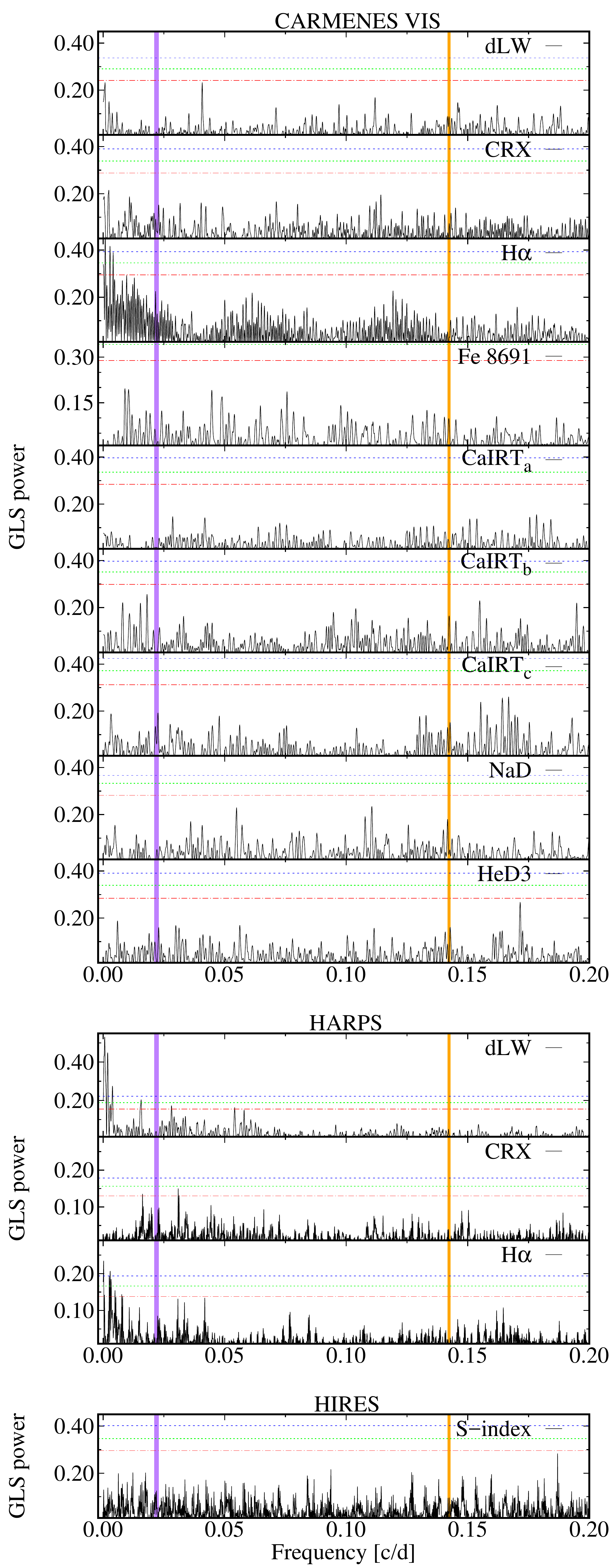}
     \caption[]{GLS periodograms of CARMENES VIS, HARPS, and HIRES activity indicators for Gl\,393. Horizontal FAP lines and vertical signal indications are the same as in Fig.~\ref{fig:RV_GLS_Gl393}.}
     \label{fig:GLS_activity_Gl393}
\end{figure}

We repeated the analysis of various activity indicators of Gl\,393 and plot their periodograms in Fig.~\ref{fig:GLS_activity_Gl393}. 
Among all indicators, only the H$\alpha$ index in CARMENES ($P=1130$\,d) and HARPS ($P=1500$\,d), as well as the HARPS-dLW ($P=4270$\,d), show significant signals. The long periods found match neither the photometric rotation period of 34.0\,d derived from K2 data nor the 7.0\,d RV signal, but could instead hint at an activity cycle of many years (which would be consistent with the temporal coherence test of Sect.~\ref{sec:coherence}; see \citealt{DiezAlonso2019} for a summary of photometric long-term cycles in M dwarfs).

\begin{figure}[]
   \centering
   \includegraphics[width=\hsize]{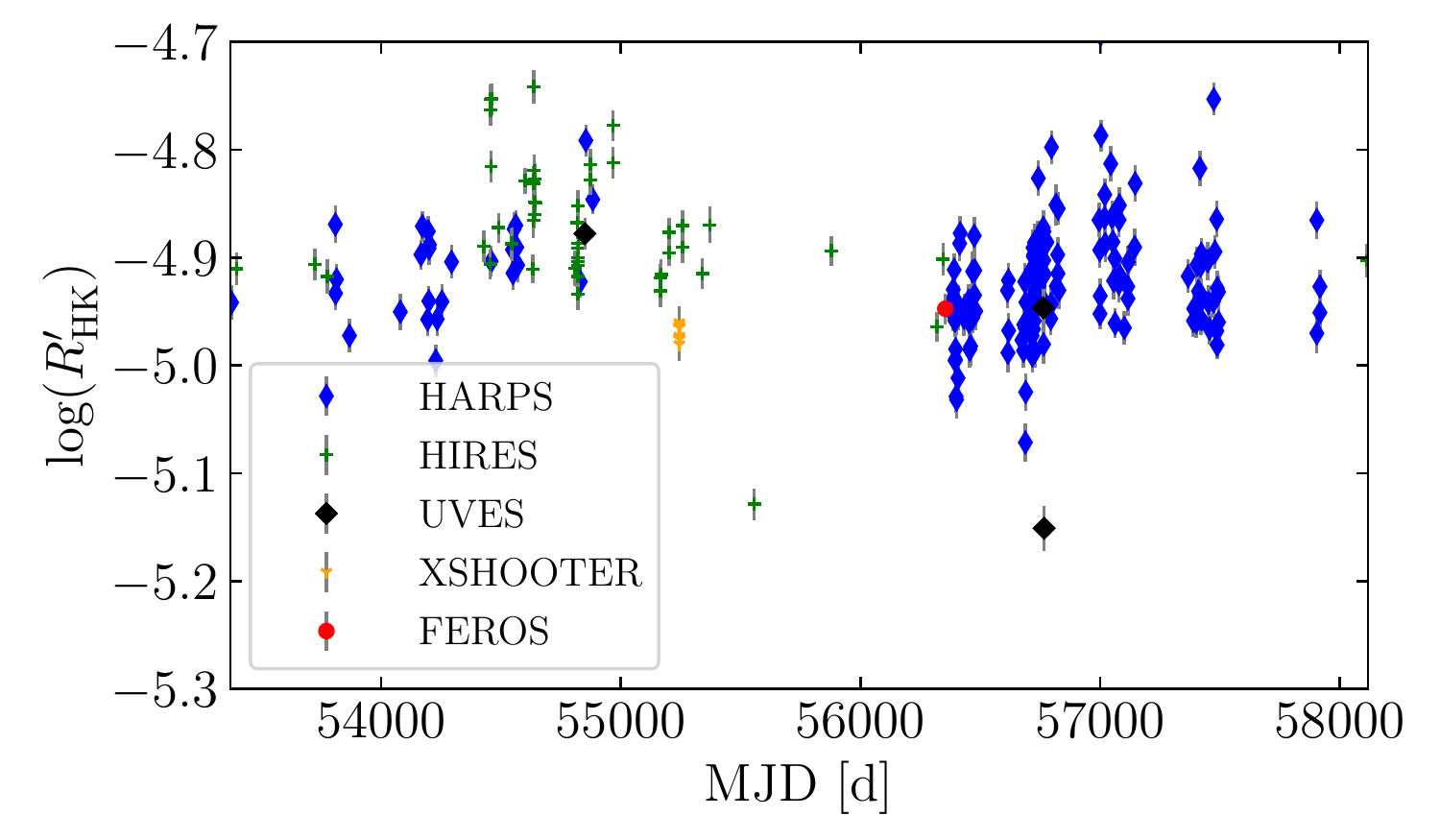}
   \includegraphics[width=\hsize]{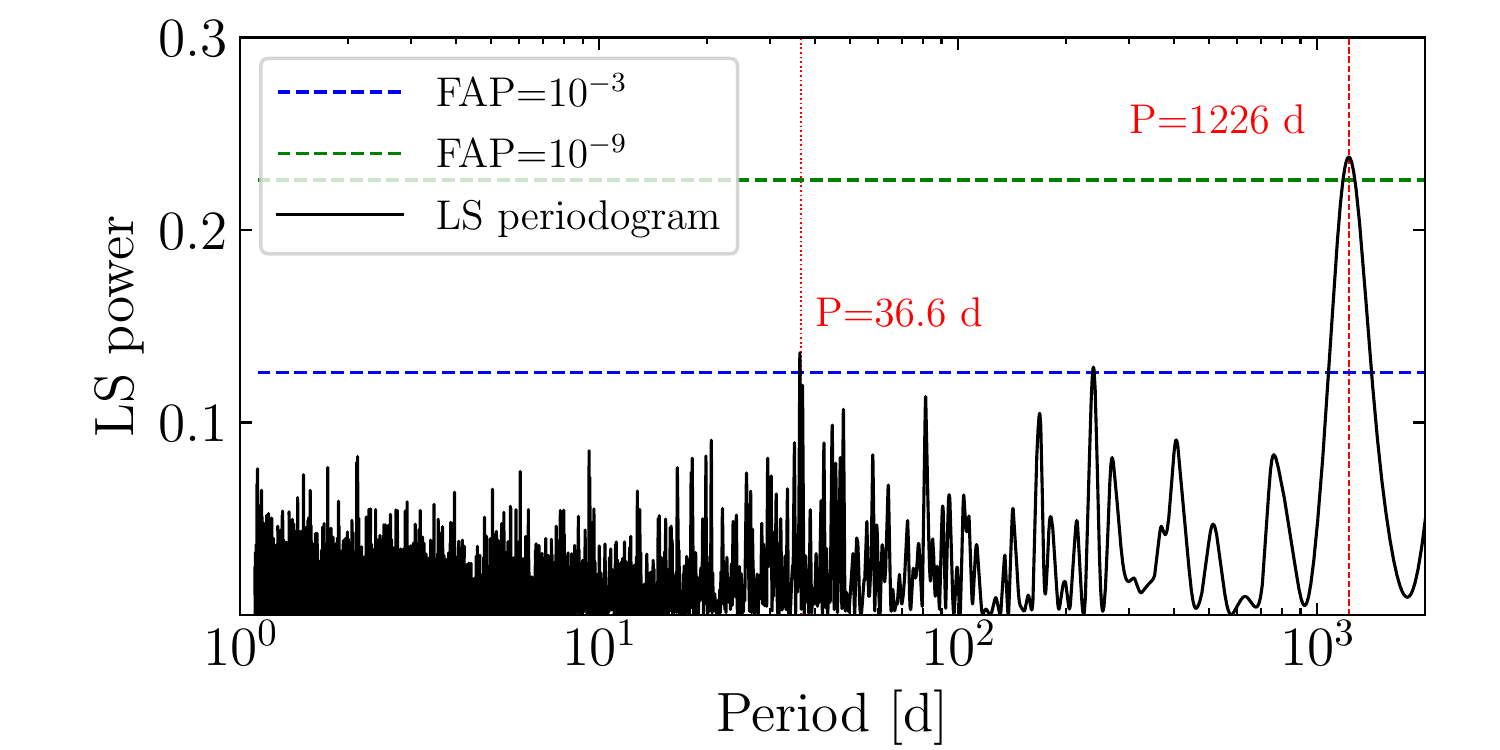}   
      \caption{{\em Top panel.}
      $R_{\rm HK}^\prime$ time series of Gl\,393.
      The source of the archival data is denoted by the colour coding.
      {\em Bottom panel.}
      Generalised Lomb-Scargle periodogram of the $R_{HK}^\prime$ time series of Gl\,393. 
      The dashed vertical lines denote the periods of the activity cycle ($P \approx$ 1226\,d) and stellar rotation ($P \approx$ 36.6\,d) determined from the $R_{HK}^\prime$ time series. 
      FAP levels of $10^{-3}$ and $10^{-9}$ are displayed as horizontal lines.}
      \label{fig:RHKp}
\end{figure}
   
Since the CARMENES spectral range does not include the Ca~{\sc ii}~H\&K lines, we carried out an analysis of all available archival data covering this spectral region, namely the HARPS and HIRES spectra used in the RV analysis of this work (Sect.~4.1.2), 1 spectrum acquired by FEROS \citep{1999Msngr..95....8K}, 3 spectra from UVES \citep{2000SPIE.4008..534D}, and 10 from XSHOOTER \citep{2011A&A...536A.105V}. 
In order to extract $R_{\rm HK}^\prime$ from spectra acquired by all five instruments in a uniform manner, we rectified the spectra with a grid of PHOENIX spectra \citep{2013A&A...553A...6H} by extracting a set of narrow band passes around both the H and K lines. 
The approach will be described in detail in an upcoming publication (Perdelwitz et al. {in prep.}). 
The top panel of Fig.~\ref{fig:RHKp} shows the resulting $R_{\rm HK}^\prime$ time series.
The smaller periodicity in the bottom panel of Fig.~\ref{fig:RHKp}, with a period of $P = 36.6\pm1.9$\,d, where the error was determined via a Monte Carlo approach, is in agreement with the stellar rotation period derived from {\em Kepler} K2 photometry (Section~3.2).

\subsection{Gaussian process analysis of RV data} \label{sec:combined_analysis}

\subsubsection{G\,264--012}

\begin{figure} 
     \includegraphics[width=\linewidth]{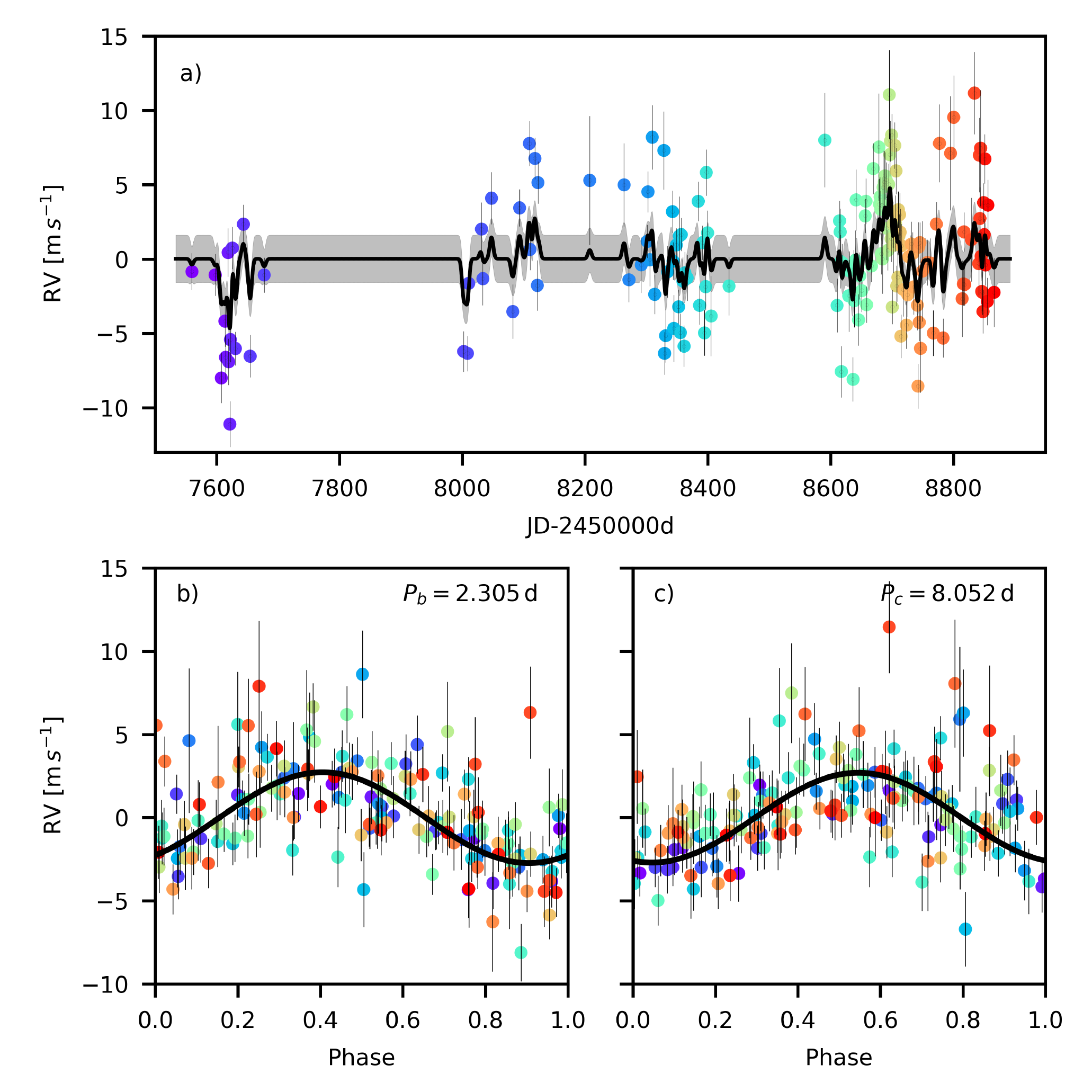} 
     \caption[]{G\,264--012. {\it Panel~a)}: GP fit (black solid line) and 1-sigma uncertainties (grey area) to CARMENES VIS RVs (points colour-coded by JD). {\it Panels~b), c)} show the phase-folded RVs after subtracting activity of the two planetary signals. Keplerian models (black solid lines) are overplotted.}
     \label{fig:Keplerians_G264} 
\end{figure} 

All indicators gathered (Sect~\ref{sec:coherence} to Sect~\ref{sec:activity_indicators}) point towards activity as origin for the RV signal found at a period of $P_{d}=92.70$\,d in G\,264--012.
There is, however, no indicator contradicting the planetary nature of the other two signals,  $P_{b}=2.30$\,d and $P_{c}=8.05$\,d.
In order to accurately determine their orbital parameters and masses, we account for stellar activity in our models. We choose a model with two circular Keplerian orbits (no evidence for eccentricity was found in these two very-short period planets, whose orbits should be almost certainly already circularised, see Sect~\ref{sec:signal_search}) for the planets combined with a Gaussian process (quasi-periodic kernel; see Eq.\,\ref{Eq1})  for the activity signal.

We present the fits in Fig.~\ref{fig:Keplerians_G264}, a summary of all parameters used in the model in Table~\ref{tab:RV+Phot_analysis_G264} and the MCMC posterior distribution plot in the appendix in Fig.~\ref{fig:cornerG}.
When subtracting the model from the VIS RV data we derived an rms of $2.22$\,m\,s$^{-1}$, which is consistent with the errorbars.

The model is very well constrained for the planetary parameters.
The periodicity $P$ and $\lambda$-timescale of the GP component are less clearly found.
The first has indistinguishable solutions ($\Delta \ln{\mathcal L}<5$) also for $P\sim 200$\,days, which is why we set a prior with $P<120$\,days, leading to an estimation for the rotational period of $102\pm4$\,days.
Solutions for $\lambda$ are not well constrained, showing probably a large variability of the spot lifetimes.
We find $64<\lambda<166$\,days, suggesting average spot lifetimes of 1 - 2.5\,$P_{\rm rot}$ \citep{Per21}.

\begin{table}
    \caption{\label{tab:RV+Phot_analysis_G264} G\,264--012: priors, best likelihoods, and posterior distribution uncertainties of the combined planets plus activity fit to RVs and derived planetary parameters$^a$.}
    \centering
    \setlength{\tabcolsep}{5.0pt}
    \renewcommand{\arraystretch}{1.3}
    \begin{tabular}{lccc}
        \hline \hline \noalign{\smallskip}
        Parameter & Prior & Posterior & Unit \\
        \noalign{\smallskip} \hline \noalign{\smallskip}
        \multicolumn{4}{c}{\em Planets}\\
        $P_{\rm b}$ & $\PriU(1.5,220)$ & $2.30538_{-0.00031}^{+0.00031}$ & d \\
        $K_{\rm b}$ & $\PriU(0,5\,\sigma_{\rm RV})$ & $2.72_{-0.29}^{+0.28}$ & m\,s$^{-1}$ \\
        $t_{c,{\rm b}}-2458000$ & $\PriU(0,220)$ & $3.235_{-0.073}^{+0.069}$ & d \\
        \noalign{\medskip}
        $P_{\rm c}$ & $\PriU(1.5,220)$ & $8.0518_{-0.0034}^{+0.0034}$ & d \\
        $K_{\rm c}$ & $\PriU(0,5\,\sigma_{\rm RV})$ & $2.69_{-0.30}^{+0.31}$ & m\,s$^{-1}$ \\
        $t_{c,{\rm c}}-2458000$ & $\PriU(0,220)$ & $4.40_{-0.23}^{+0.24}$ & d \\
        \noalign{\smallskip}
        \multicolumn{4}{c}{\em GP kernel hyperparameters}\\
        $h$ & $\PriU(0,5\,\sigma_{\rm RV})$ & $1.59_{-0.36}^{+0.36}$ & m\,s$^{-1}$ \\
        $P$ & $\PriU(0,120)$ & $102.2_{-4.4}^{+4.2}$ & d \\
        $\lambda$ & $\PriU(0,\sim 2800)$ & $96_{-32}^{+70}$ & d \\        
        $w$ & $\PriU(0,1)$ & $0.271_{-0.089}^{+0.108}$ &  \\        
       \noalign{\smallskip}
        \multicolumn{4}{c}{\em Instruments} \\
        $\sigma$ & $\PriU(0 ,3\,\sigma_{\rm RV})$ & $0.48_{-0.36}^{+0.36}$ & m\,s$^{-1}$ \\
        $\mu$ & $\PriU(\mu_1,\mu_2 )$ & $0.02_{-0.54}^{+0.53}$ & m\,s$^{-1}$ \\
        \noalign{\smallskip}
        \multicolumn{4}{c}{\em Derived}\\
        $a_{\rm b}$ & & $0.02279_{-0.00061}^{+0.00061}$ & au\\
        $M_{\rm b} \sin{i}$ & & $2.50_{-0.30}^{+0.29}$ & $M_\oplus$ \\
        $S_{\rm b}$ & & $20.5_{-1.1}^{+1.1}$ & $S_{\oplus}$ \\
        $T_{\rm{eq,b}}$ & & $587_{-16}^{+16}$  & K \\
        \noalign{\medskip}
        $a_{\rm c}$ & & $0.0525_{-0.0014}^{+0.0014}$ & au\\
        $M_{\rm c} \sin{i}$ & & $3.75_{-0.47}^{+0.48}$ & $M_\oplus$ \\
        $S_{\rm c}$ & & $3.87_{-0.21}^{+0.21}$ & $S_{\oplus}$ \\
        $T_{\rm{eq,c}}$ & & $387_{-11}^{+11}$  & K \\
        \noalign{\smallskip} \hline 
	\end{tabular}
	\tablefoot{$^a$ $\mu_1 = \overline{RV} - 3\,\sigma_{\rm RV}$, and $\mu_2 = \overline{RV} + 3\,\sigma_{\rm RV}$, with $\overline{RV}$ and $\sigma_{\rm RV}$, as the average and the rms of the RV time series, respectively.
    }

\end{table}

\subsubsection{Gl\,393}
\label{sec:combined.Gl393}

\begin{figure} 
     \includegraphics[width=\linewidth]{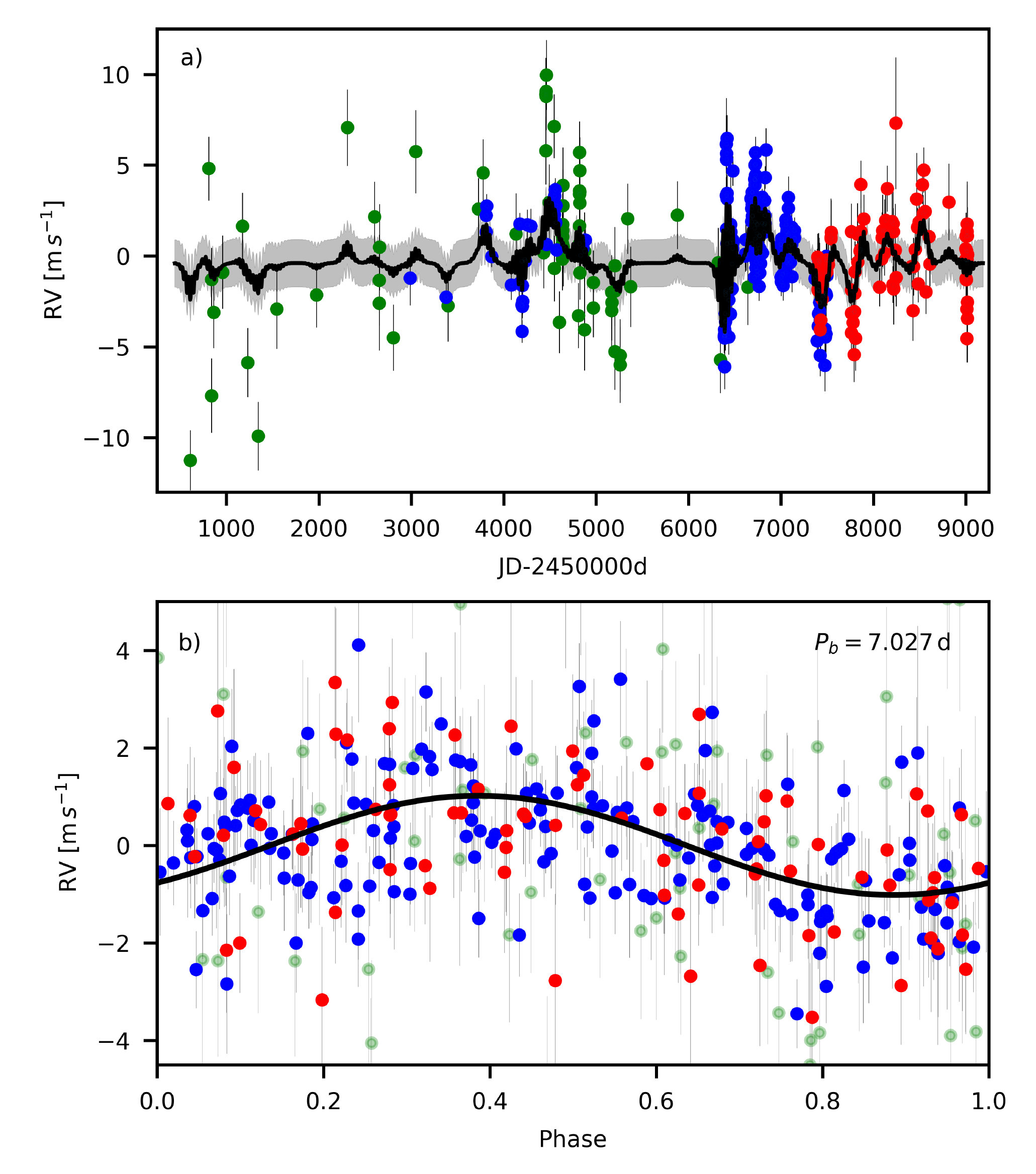}
     \caption[]{Gl\,393. {\it Panel~a)}: GP fit (black solid line) and 1-sigma uncertainties (grey area) to the RV data of HIRES (green points), HARPS (blue points) and CARMENES VIS channel (red squares). {\it Panel~b)}: activity subtracted RVs, phase-folded to the $7.03$\,d period. The Keplerian model (black solid line) is overplotted.}
     \label{fig:Keplerians_Gl393} 
\end{figure}

Gl\,393 has a rotational period of 34.0\,d as derived from K2 photometry and in agreement with the periodicity of the $\ion{Ca}{ii}$\,H\&K variability. Spot patterns seem to evolve rapidly, giving rise to a complex, activity driven RV variability. To derive the parameters of the planetary signal discovered in the RV data, activity modelling is crucial. We proceeded in the same manner as for G\,264--012 and fitted a circular Keplerian model (no eccentric orbit solution was found to be significant from our analysis in Sect.~\ref{sec:signal_search}) combined with a GP using a quasi-periodic kernel which accounted best for the evolving activity signal. 
The resulting fits to the data are presented in Fig.~\ref{fig:Keplerians_Gl393} and the parameters used when deriving the models are summarised in Table~\ref{tab:RV+Phot_analysis_Gl393}.
The MCMC posterior distribution plot is shown in the appendix in Fig.~\ref{fig:cornerGl393}.
When subtracting the model from the data we find an rms of $3.32$\,m\,s$^{-1}$ for HIRES, $1.19$\,m\,s$^{-1}$ for HARPS and $1.50$\,m\,s$^{-1}$ for CARMENES VIS; all are consistent with the respective error estimates in each data set.
All parameters are very well constrained, with the exception of the periodicity $P$ which, resembling the results of the periodogram analysis, tends to vary between 30 and 50\,days, and with a maximum likelihood for P=35.0\,days.
We calculate an average spot lifetime of 2 - 3\,$P_{\rm rot}$ from the $\lambda$-hyperparameter.

\begin{table}
    \caption{\label{tab:RV+Phot_analysis_Gl393} Gl\,393: priors, best likelihood and average posterior parameter values of the planet plus activity fit to RVs, and derived planetary parameters.}
    \centering
    \setlength{\tabcolsep}{5.0pt}
    \renewcommand{\arraystretch}{1.3}
    \begin{tabular}{@{}lccl@{}}
        \hline
        \hline
        \noalign{\smallskip}
        Parameter & Prior & Posterior & Unit \\
        \noalign{\smallskip}
        \hline
        \noalign{\smallskip}
        \multicolumn{4}{c}{\em Planets}\\
        $P_{\rm b}$ & $\PriU(0,100)$ & $7.02679_{-0.00085}^{+0.00082}$ & d \\
        $K_{\rm b}$ & $\PriU(0,5\,\sigma_{\rm RV})$ & $1.01_{-0.14}^{+0.14}$ & m\,s$^{-1}$ \\
        $t_{c,{\rm b}}-2458000$ & $\PriU(0,100)$ & $2.72_{-0.21}^{+0.21}$ & d \\
        \noalign{\smallskip}
        \multicolumn{4}{c}{\em GP kernel hyperparameters}\\
        $h$ & $\PriU(0,5\,\sigma_{\rm RV})$ & $1.30_{-0.21}^{+0.22}$ & m\,s$^{-1}$ \\
        $P$ & $\PriU(0,220)$ & $35.0_{-4.3}^{+7.6}$ & d \\
        $\lambda$ & $\PriU(0,\sim 19000)$ & $54.3_{-11.1}^{+13.6}$ & d \\        
        $w$ & $\PriU(0,1)$ & $0.73_{-0.13}^{+0.13}$ &  \\        
       \noalign{\smallskip}
        \multicolumn{4}{c}{\em Instruments} \\
        $\sigma_{\text{HIRES}}$ & $\PriU(0,3\,\sigma_{\rm RV}))$ & $1.927_{-0.474}^{+0.406}$ & m\,s$^{-1}$ \\
        $\mu_{\text{HIRES}}$ & $\PriU(\mu_1,\mu_2)$ & $-0.39_{-0.52}^{+0.52}$ & m\,s$^{-1}$ \\
        $\sigma_{\text{HARPS}}$ & $\PriU(0,3\,\sigma_{\rm RV})$ & $0.004_{-0.071}^{+0.128}$ & m\,s$^{-1}$ \\
        $\mu_{\text{HARPS}}$ & $\PriU(\mu_1,\mu_2)$ & $-4.25_{-0.42}^{+0.41}$ & m\,s$^{-1}$ \\
        $\sigma_{\text{C.VIS}}$ & $\PriU(0,3\,\sigma_{\rm RV})$ & $0.015_{-0.143}^{+0.246}$ & m\,s$^{-1}$ \\
        $\mu_{\text{C.VIS}}$ & $\PriU(\mu_1,\mu_2)$ & $0.23_{-0.52}^{+0.52}$ & m\,s$^{-1}$ \\
        \noalign{\smallskip}
        \multicolumn{4}{c}{\em Derived}\\
        $a_{\rm b}$ & & $0.05402_{-0.00072}^{+0.00072}$ & au\\
        $M_{\rm b} \sin{i}$ & & $1.71_{-0.24}^{+0.24}$ & $M_\oplus$ \\
        $S_{\rm b}$ & & $9.21_{-0.31}^{+0.31}$ & $S_{\oplus}$ \\
        $T_{\rm{eq,b}}$ & & $485_{-11}^{+11}$  & K \\
        \noalign{\smallskip}
        \hline 
\end{tabular}
\end{table}

\section{Discussion}\label{sec:discussion}

\subsection{Comparison with other surveys}

\begin{figure*}
     \includegraphics[width=\linewidth]{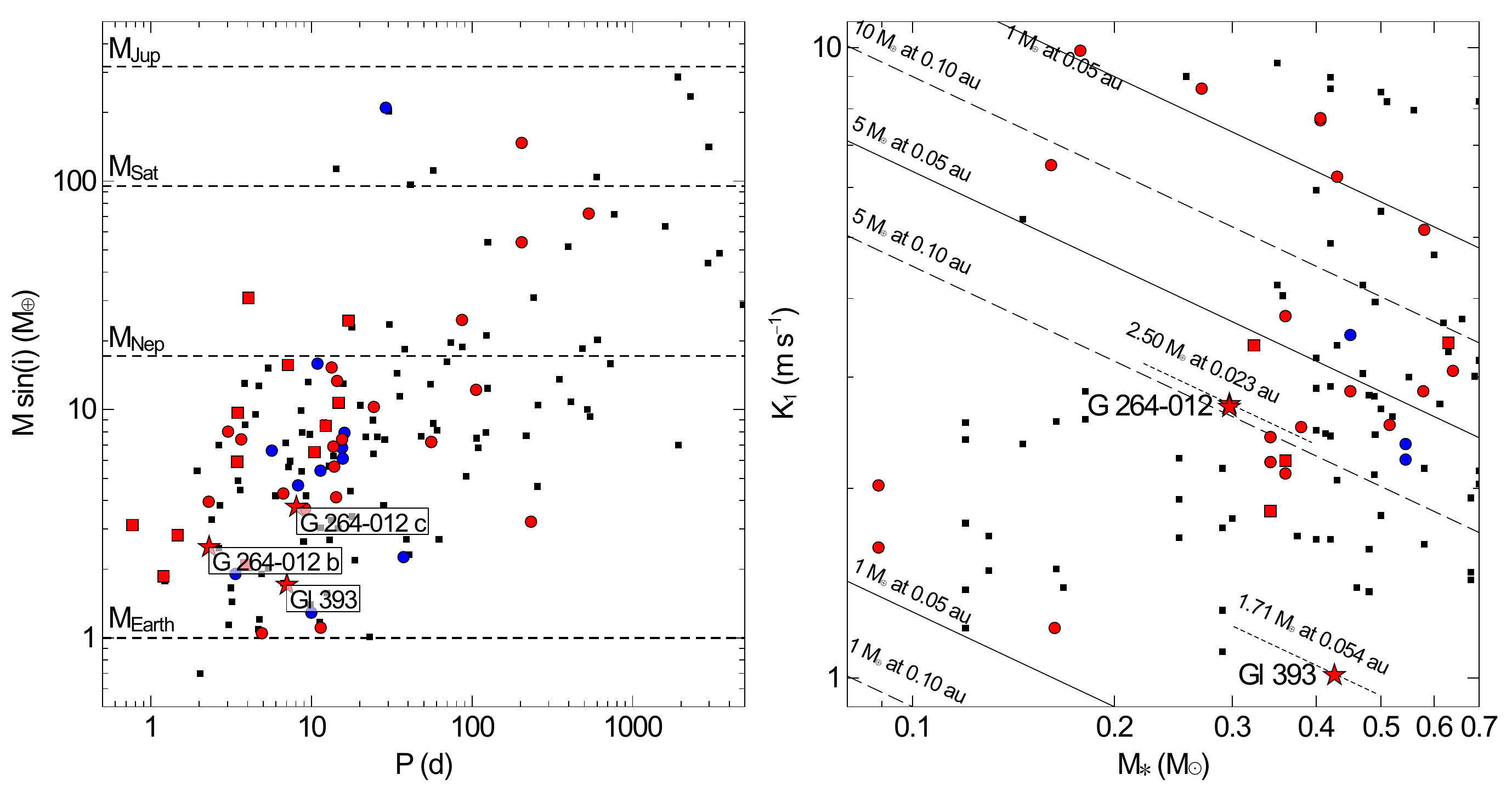}
     \caption[]{Parameters of all known exoplanets around M dwarfs. {\it Left panel}: Minimum planet mass versus orbital period. {\it Right panel}: RV curve semi-amplitude against host stellar mass. Black squares represent exoplanet parameters for those listed in \url{http//exoplanets.eu} on 01 February 2021 and cyan small circles represent TESS detections. Red symbols show planets detected by the CARMENES survey (solid circles) or by its follow-up of {\em TESS} detections (solid squares). The three planet candidates presented in this work are shown as red stars.}
     \label{fig:semi-amp} 
\end{figure*}

As discussed in Sect.~\ref{sec:star}, G\,264--012 and Gl\,393 were catalogued before as potential targets for RV surveys for exoplanets.
However, only Gl\,393 was extensively monitored by HARPS \citep{Bonf13}, HIRES \citep{Butl17}, and ESPRESSO \citep{Hojj19}. In contrast, G\,264--012 was not observed with high-precision RVs until now.
The exoplanet Gl\,393\,b may have passed unnoticed due to its low mass and the strong impact of the magnetic activity of the star on the RV variability. Fig.~\ref{fig:semi-amp} shows that this planet produces the smallest RV semi-amplitude of any of all detected planets around M dwarfs.

In their catalogue of chromospheric activity and jitter measurements of 2630 stars in the California Planet Search with HIRES, \cite{IsaacsonFischer2010} found Gl\,393 to be active and RV variable with a jitter of 3.174\,m\,s$^{-1}$.
\cite{Gran20}, using HARPS data, also found its RV curve to be dominated by spots. Though they used the correlation between RVs and the bisector velocity span (BVS; or the velocity span covered by the bisector of the cross-correlation function) to correct the RVs from the signal caused by spots, as in \cite{Melo07}, they still did not detect the planet.

At that time, the star showed an RV amplitude of 17.7\,m\,s$^{-1}$ (BVS of 24.5\,m\,s$^{-1}$) and 1.2 and 3.6\,m\,s$^{-1}$ in the mean RV uncertainty and rms, respectively, whereas the CARMENES RV data acquired during the latter epochs has values of 1.5 and 2.3\,m\,s$^{-1}$ in the mean RV uncertainty and rms.
The combined GP+Keplerian model takes a big leap in evidence, showing the much lower activity level of the star when the CARMENES data were obtained as compared with previous seasons.
This is confirmed by the fact that the rotational period signal in our RVs loses power over the seasons.

The HARPS non-detection probably happened because their data were acquired during epochs when the star was dominated by activity.
This shows that the detection of exoplanets depends on the activity level of the star which, in turn, depends on when the planet is observed within its activity cycle.
Therefore, planets could be detected in moderately active M dwarfs if observed during their magnetic cycle minima or, at least, away from their maxima, when activity signals dominate their time series.

\subsection{The three planets in context}

Being M4.0V and M2.0V spectral-type stars, G\,264--012 and Gl\,393 have liquid-water habitable zones with optimistic inner edges at 0.087 and 0.136\,au from their stars, respectively \citep{kopp13}.
The three planets in these two systems orbit closer to their stars, at 0.023 and 0.053\,au, for planets G\,264--012~b and c, and at 0.054\,au for Gl\,393~b, than these optimistic limits.
At the distances from their stars given by their semi-major axes, all three planets receive larger fluxes than that received at Earth from the Sun, reaching higher equilibrium temperature, assuming zero Bond albedo (see Tables~\ref{tab:RV+Phot_analysis_G264} and \ref{tab:RV+Phot_analysis_Gl393}).

As can be seen in Fig.~\ref{fig:semi-amp}, all three planets lie on the Earth-mass region of the parameter space of the mass versus period diagram. 
The ratios of the minimum mass of the planets to that of the star for both systems is similar, with values of 6.32 and 1.21 times $10^{-5}$ for G\,264--012 and Gl\,393, respectively.

\subsection{Formation scenarios}

The three planets presented in this paper are close-in (super-) Earths and representations of the lower-mass end of their population. 
Previous findings of similar planets \citep[e.g.][]{luqu19}, as well as theoretical works (Burn et~al., accepted), suggest that the planets we detected are a frequent outcome of planet formation around M dwarfs.
The solar metallicities of the host stars are thought to be favourable for the formation of such inner low-mass planet systems \citep{Schl20}.

Planets of this type are thought to form via core accretion \citep{Perr74, Mizu78, Mizu80}, which relies on the accretion of solid material via planetesimals and smaller particles. 
The accretion of these smaller particles, or `pebbles', typically millimetre- to centimetre-sized bodies, was shown to be very efficient in growing planetary cores \citep{Orme10,Lamb12}. 
Since fresh supply of such material is ensured by a radial flux in the protoplanetary disc \citep{Lamb14}, there must be some mechanism at work that stops the accretion to ultimately yield the low masses that we measure. 
Plausible mechanisms for such a cut-off are self-isolation through perturbations of the surrounding disc \citep{Morb12,Lamb14b} or the emergence of a hypothetical massive outer companion \citep[e.g.][]{Orme17}.

However, M-dwarf planets similar to G\,264--012\,b and c and Gl\,393\,b are also well reproduced by models that grow planetary cores only via accretion of planetesimals and do not take into account pebble accretion \citep[e.g.][]{Emse20}. This variant requires no assumptions about a pebble blocking mechanism.

Both models typically assume the formation of planets further out with a subsequent migration, but in-situ formation of super-Earths may be possible under certain conditions, such as super-solar metallicity of the host star \citep[e.g.][]{Schl14}.
Our discovery reinforces the emerging paradigm that the formation of compact systems of low-mass planets around M dwarfs is the rule rather than the exception.

\subsection{Planet detectability}

\begin{figure}
     \includegraphics[width=\linewidth]{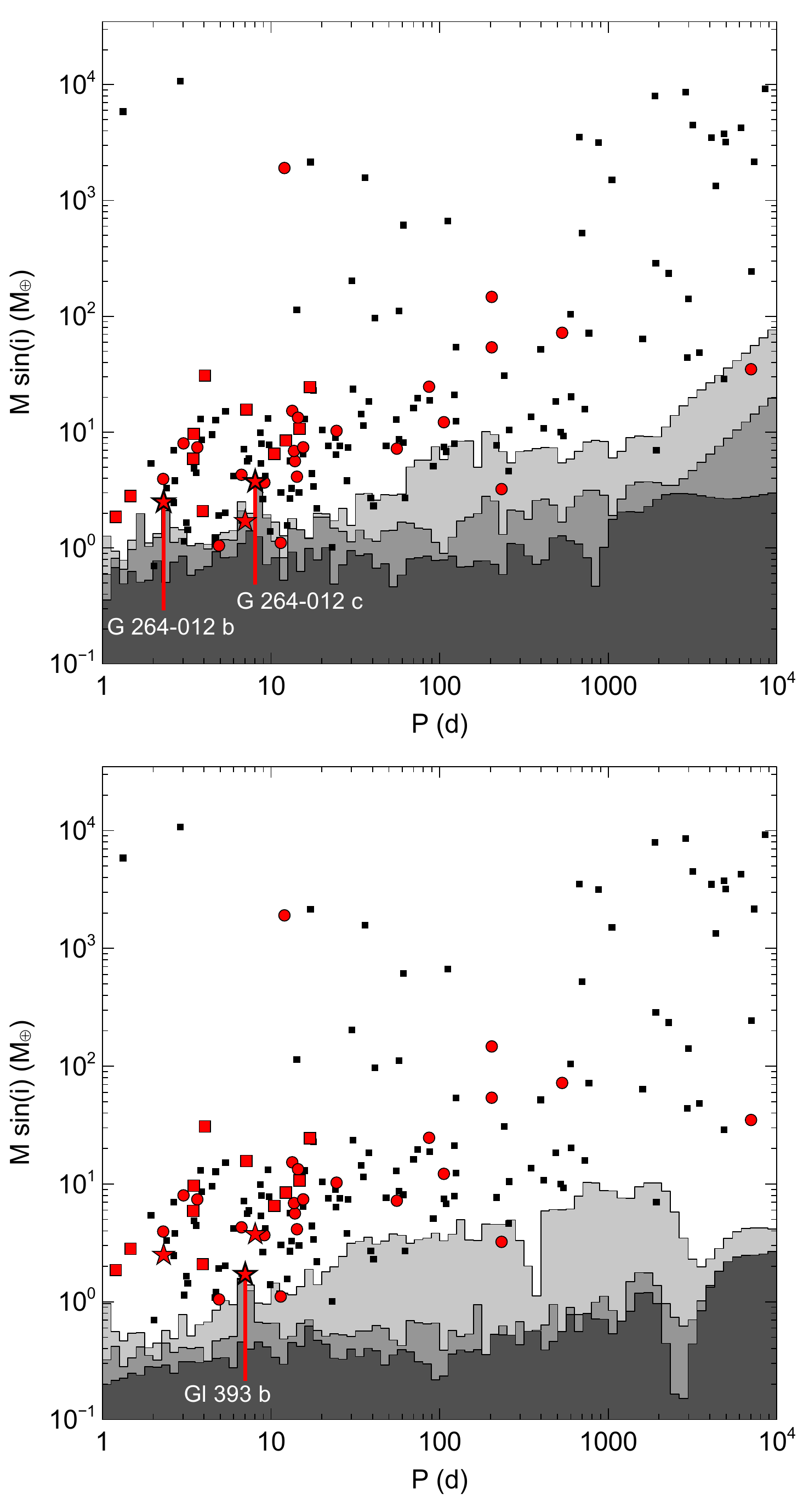}
     \caption[]{Maximum $M\sin{i}$ value compatible with the RV measurements as a function of prospective orbital period. {\it Upper panel}: for G\,264--012 RV curve.  {\it Lower panel}. Gl\,393 RV curve. From light to darker grey, detectability from original RV data (no modelling of planets or activity), data corrected for activity with GP, and data corrected from both, activity and planet signals. Symbols are the same as in Fig.~\ref{fig:semi-amp} with the planets in each of the two systems labelled only with b, c for G\,264--012 and b for Gl\,393.}
     \label{fig:detectability} 
\end{figure} 

Though the data at hand are insufficient to unequivocally distinguish between the different formation models, they provide constraints for the existence of larger outer planets in these two systems. To this aim, we estimated, for each of our two stars, the maximum $M\sin{i}$ value compatible with the RV measurements as a function of prospective orbital period.
First, we verified that the RVs of our two stars did not show long-term trends, which would have needed correction, and fitted the data to a circular orbit using a partially-linearised, least-squares fitting procedure\footnote{{\tt Minuit}: \url{https://iminuit.readthedocs.io/en/stable/about.html}} \citep{Roos75} as a function of different orbital periods $P$.
For each prospective $P$, we then determined the best-fit semi-amplitude $K(P)$ and computed a planet mass value $M\sin{i_{\rm{max}}}(P) = M_m(P)$ using the mass of the star. This mass is the maximum a planet can have to be non-detectable in our RV time series, considering their respective noise characteristics.

Figure~\ref{fig:detectability} shows $M_m(P)$ as a function of the orbital period for G\,264--012 and Gl\,393. Both stars have good quality observations, with a large number of data points.
Nonetheless, the much longer baseline and larger number of RVs acquired for Gl\,393, with three different instruments, make this data set sensitive to planets with masses as low as $M\sin{i} \approx 0.3$\,M$_{\oplus}$ for $P < 3$\,d and below 1.0\,M$_{\oplus}$ for periods up to 2000\,d.
The sensitivity of G\,264--012 RV data set is somewhat less due to the shorter time span of its time series acquired with CARMENES only (compare dark-grey areas for both stars in Fig.~\ref{fig:detectability}).
For this star, the change of slope of the sensitivity for periods beyond the length of the data set is apparent at around 1400\,d.
The figure also shows the differences in detectability of our raw data (no cleaning of planets or activity) compared with data cleaned from activity using GP, and fully cleaned data from planetary and activity signals.
By removing the activity, the peak corresponding to Gl\,393\,b becomes more significant in the detectability maps.
After removing the RV variability produced by the activity and the planets, the remaining detectability functions show a higher sensitivity at all periods.

These RVs disprove the presence in the system of planets larger than $\sim 1.0$\,M$_{\oplus}$ around G\,264--012 up to periods of 1000\,d.
Gl\,393 does not show signs of any other planet with mass above $\sim 0.5$\,M$_{\oplus}$ up to periods of 100\,d and up to $\sim 1.0$\,M$_{\oplus}$ in the 100--1000\,d period range.
In Fig.~\ref{fig:detectability}, we also plot all known exoplanets around M dwarfs, showing that our RVs are of enough quality to show the signal of any of those planets if they had been orbiting our systems.

\subsection{Dynamical stability of G264-012}

In this subsection, we seek to test the stability of the G264-012 system by varying the parameters that remained unconstrained after the global model: the inclination of the system from the observer point of view, which directly affects the real masses of the planets, and their eccentricities. We note that the natural trend of planetary systems is to reside in a nearly co-planar configuration; hence, we assumed planets b and c as being co-planar. This choice was further motivated by the results yielded by Kepler and the HARPS survey, which suggest that the mutual inclinations of multi-planet systems are of the order $\leq$10~deg \citep[see e.g.][]{fabrycky2014}.

To this end, we made use of the Mean Exponential Growth factor of Nearby Orbits (MEGNO) chaos index, $Y(t)$ \citep{cincottasimo1999,cincottasimo2000,cincotta2003}. MEGNO is extensively used to test the stability of many planetary systems \cite[e.g.][]{jenkins2019,pozuelos2020,demory2020}. In particular, its time-averaged mean value, $\langle Y(t) \rangle$, amplifies any stochastic behaviour, allowing for the detection of hyperbolic regions during the integration time. Hence, $\langle Y(t) \rangle$ allows us to distinguish between chaotic and stable trajectories: if $\langle Y(t) \rangle \rightarrow \infty$ for $t\rightarrow \infty$, the system is chaotic; while if $\langle Y(t) \rangle \rightarrow 2$ for $t\rightarrow \infty$, the motion is stable. We used the MEGNO implementation within the N-body integrator \texttt{REBOUND} \citep{rein2012}, which makes use of the Wisdom-Holman \texttt{WHFast} code \citep{rein2015}.

First, we evaluated the inclination of the system 1000 times, investigating values from 10 to 90~deg. Our choice to limit the inclination to 10~deg is twofold; first, randomly oriented systems disfavour small values of inclination \cite[see e.g.][]{dreizler2020}; and second, orbital inclinations close to zero (face on) disfavour the detection of planetary systems via RVs. From our global analysis, the results hint that planets b and c reside in circular orbits; hence, in this case we considered zero eccentricity for both planets. We found that the system tolerated the full range of inclinations explored with a $\Delta \langle Y(t) \rangle$= 2.0 $-$ $\langle Y(t) \rangle <10^{-3}$. This dynamical robustness impeded strong constraints on the system's inclination, and consequently on the planetary masses. While our global model favoured circular orbits, mutual interactions between planets b and c may still produce orbital excitations which introduce some level of eccentricities. Then, to estimate their maximum values without disrupting the system's stability, we conducted a second suite of simulations by allowing the planets to have eccentric orbits at different system's inclinations. We built stability maps $e_{b}$--$e_{c}$ for discrete sample of inclinations of 90, 70, 50, 30 and 10~deg, and eccentricities in the range 0.0--0.3. We took 10 values from this range, meaning that the size of each stability map was 10$\times$10~pixels. We found that, in the case of maximum system inclination (90~deg), that is, minimum planet masses, both planets must have eccentricities $<0.3$ to ensure the stability of the system. In the case of minimum inclination (10~deg), or when the masses of the planets are closer to their maximum values, the eccentricities must be $<0.1$. This means that, while circular orbits represent the most plausible architecture of the system with the data at hand, planets b and c may tolerate a certain level of eccentricity, but in all cases the eccentricity are very likely to be below 0.3. In all our simulations, the integration time was set to $10^{6}$ times the orbital period of the outermost planet, c, and the integration time-step was 5\% the orbital period of the innermost one, b.

\section{Conclusions}\label{sec:conclusions}

We studied two systems within the CARMENES survey of exoplanets around M dwarfs.
Our analyses of CARMENES-only data for G\,264--012 and of CARMENES, HARPS, and HIRES RV data for Gl\,393 suggest that these two stars are orbited by two and one super-Earth planets, respectively.
The planets around G\,264--012 have minimum masses of 2.5 and 3.8\,M$_{\oplus}$ with orbital periods of 2.3 and 8.1\,d, corresponding to semi-major axes of 0.02 and 0.05\,au. 
We calculated the equilibrium temperature of G\,264--012~b and c to be around 590 and 380~K, respectively.
Gl\,393 is orbited by a terrestrial planet with an orbital period of 7\,d. Gl\,393\,b has a minimum mass of roughly 1.7\,M$_{\oplus}$, and is orbiting its host star at a distance of approximately 0.054\,au.
Its equilibrium temperature is estimated to be of around 480\,K.

Our minimum mass and temperature estimates put the planets into the family of hot Earths and Super-Earths.
Since they receive total radiative fluxes larger than that of Earth, the planets are too close to the host star for the possible existence of liquid water on their surfaces.

The exoplanets in these two systems fulfil all the characteristics to be perfect targets for radio observations to test the star-planet interaction scenario.
This hypothesis was recently put forward to explain the detection of coherent circularly polarised radio emission in GJ\,1151 \citep{Ved20}.
This type of emission was even found to be modulated with the orbital period of the planet in the only other instance of such detections, \object{Proxima Centauri} \citep{Pere21}.
For the two systems described in this work, the computed flux density expected from the star-planet interaction is within the reach of current radio interferometers.

The detection of auroral radio emission from the stars periodically modulated with the orbital period of the planet, as shown for Proxima\,b \citep{Pere21}, would be clear evidence of star-planet interaction. In the case of GJ\,1151, a RV signal ascribed to a possible candidate planet at an orbital period suggested by the previously detected radio emission was found by \cite{Mah21} but later refuted by \cite{Per21b}.
The detection of such emission, with the required characteristics, in the two systems presented in this work would independently confirm the presence of their exoplanets and help validate radio observations as a completely new and independent planet-detection technique.

\begin{acknowledgements} 

CARMENES is an instrument for the Centro Astron\'omico Hispano-Alem\'an (CAHA) at Calar Alto (Almer\'{\i}a, Spain), operated jointly by the Junta de Andaluc\'ia and the Instituto de Astrof\'isica de Andaluc\'ia (CSIC).
  
CARMENES was funded by the Max-Planck-Gesellschaft (MPG), the Consejo Superior de Investigaciones Cient\'{\i}ficas (CSIC), the Ministerio de Econom\'ia y Competitividad (MINECO) and the European Regional Development Fund (ERDF) through projects FICTS-2011-02, ICTS-2017-07-CAHA-4, and CAHA16-CE-3978, and the members of the CARMENES Consortium (Max-Planck-Institut f\"ur Astronomie, Instituto de Astrof\'{\i}sica de Andaluc\'{\i}a, Landessternwarte K\"onigstuhl, Institut de Ci\`encies de l'Espai, Institut f\"ur Astrophysik G\"ottingen, Universidad Complutense de Madrid, Th\"uringer Landessternwarte Tautenburg, Instituto de Astrof\'{\i}sica de Canarias, Hamburger Sternwarte, Centro de Astrobiolog\'{\i}a and Centro Astron\'omico Hispano-Alem\'an), with additional contributions by the MINECO, the Deutsche Forschungsgemeinschaft through the Major Research Instrumentation Programme and Research Unit FOR2544 ``Blue Planets around Red Stars'',  the Klaus Tschira Stiftung, the states of Baden-W\"urttemberg and Niedersachsen, and by the Junta de Andaluc\'{\i}a.

We acknowledge financial support from the Agencia Estatal de Investigaci\'on of the Ministerio de Ciencia, Innovaci\'on y Universidades and the ERDF through projects    
  PID2019-109522GB-C5[1:4]/AEI/10.13039/501100011033,	
  PGC2018-098153-B-C33,		
  AYA2017-89637-R           
and the Centre of Excellence ``Severo Ochoa'' and ``Mar\'ia de Maeztu'' awards to the Instituto de Astrof\'isica de Canarias (SEV-2015-0548), Instituto de Astrof\'isica de Andaluc\'ia (SEV-2017-0709), and Centro de Astrobiolog\'ia (MDM-2017-0737), 
the Funda\c{c}\~ao para a Ci\^encia e a Tecnologia and the ERDF (COMPETE2020), the Programa Operacional Competitividade e Internacionaliza\c{c}\~ao (UID/FIS/04434/2019, 
UIDB/04434/2020, 
UIDP/04434/2020, PTDC/FIS-AST/[32113,28953,28987]/2017 \& POCI-01-0145-FEDER-[032113,028953,028987]),
the Generalitat de Catalunya (CERCA programme), the European Research Council under the Horizon 2020 Framework programme (ERC Advanced Grant Origins 832428),
FONDECYT (3180063), 
the Swiss National Science Foundation for supporting research with HARPS (SNSF 140649, 152721, 166227, and 184618), 
and NASA (NNX17AG24G).

This publication made use of the SIMBAD database, the Aladin sky atlas, and the VizieR catalogue access tool developed at CDS, Strasbourg Observatory, France, 
the Python libraries {\tt Matplotlib}, {\tt NumPy}, and {\tt SciPy}, the collection of software packages {\tt AstroPy} and {\tt topfplotter}, and data from the CARMENES data archive at CAB (CSIC-INTA), the ESO Science Archive Facility under request number {vperdelw-552557}, the {\em TESS} mission, obtained from the MAST data archive at the Space Telescope Science Institute (STScI), and the K2 mission.
\end{acknowledgements} 


\bibliographystyle{aa} 
\bibliography{References} 


\begin{appendix}
\onecolumn
\section{Corner plot of MCMC posterior distributions}
\begin{figure}[hbt!]
	\includegraphics[width=\linewidth]{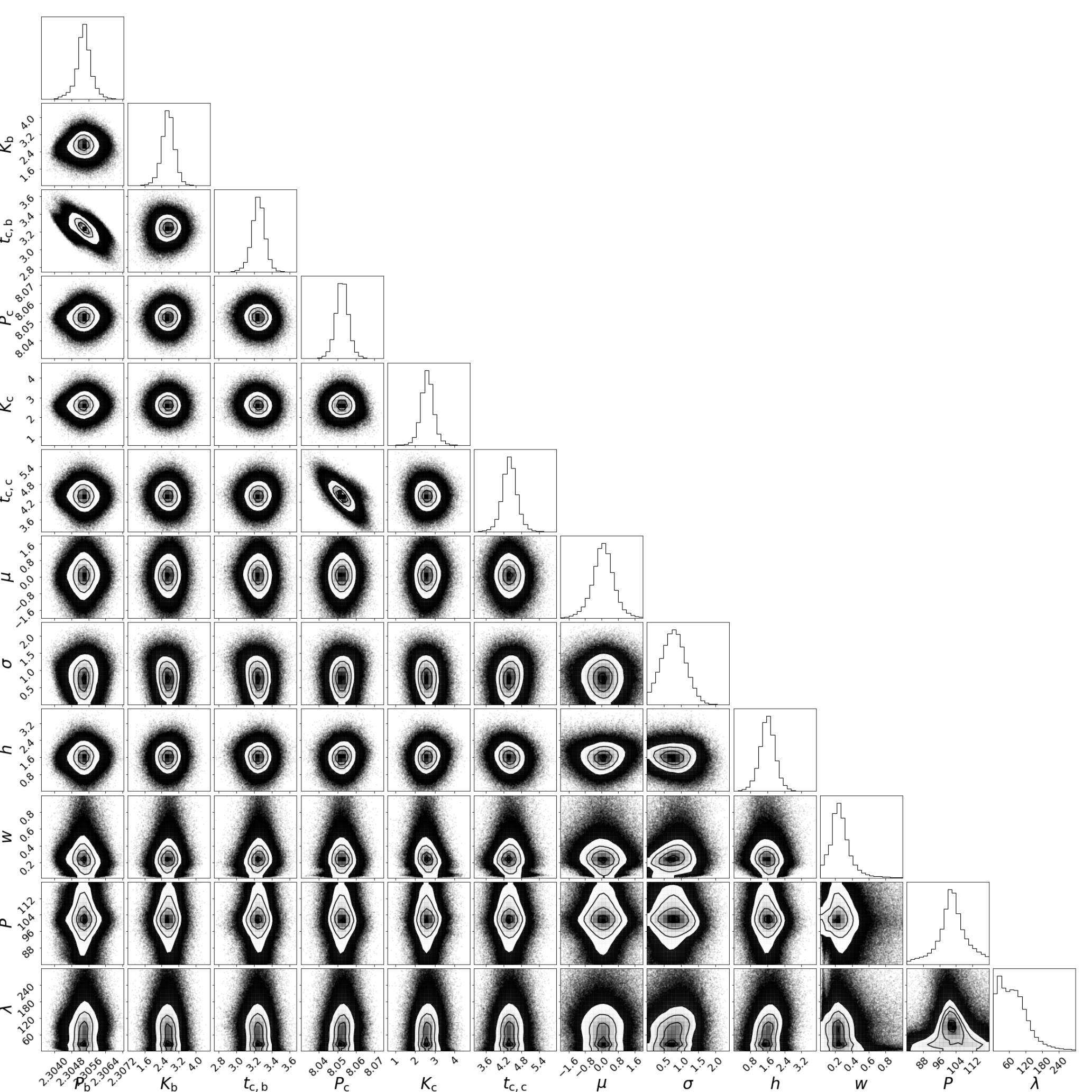}
	\caption[]{MCMC posterior distribution of the CARMENES VIS RVs for G264-012 for a model including two circular planetary orbits (parameters $K_b$, $P_b$, $T_{\rm c,b}$, $K_c$, $P_c$, and $T_{\rm c,c}$), a Gaussian Process using the {\tt george}-code and a quasi-periodic kernel (hyper-parameters $h$, $w$, $P$, and $\lambda$), offset $\mu$ and additional jitter $\sigma$. Only solutions with $\ln{\mathcal L}>-410$ are shown.}
	\label{fig:cornerG} 
\end{figure} 	

\begin{figure}
	\includegraphics[width=\linewidth]{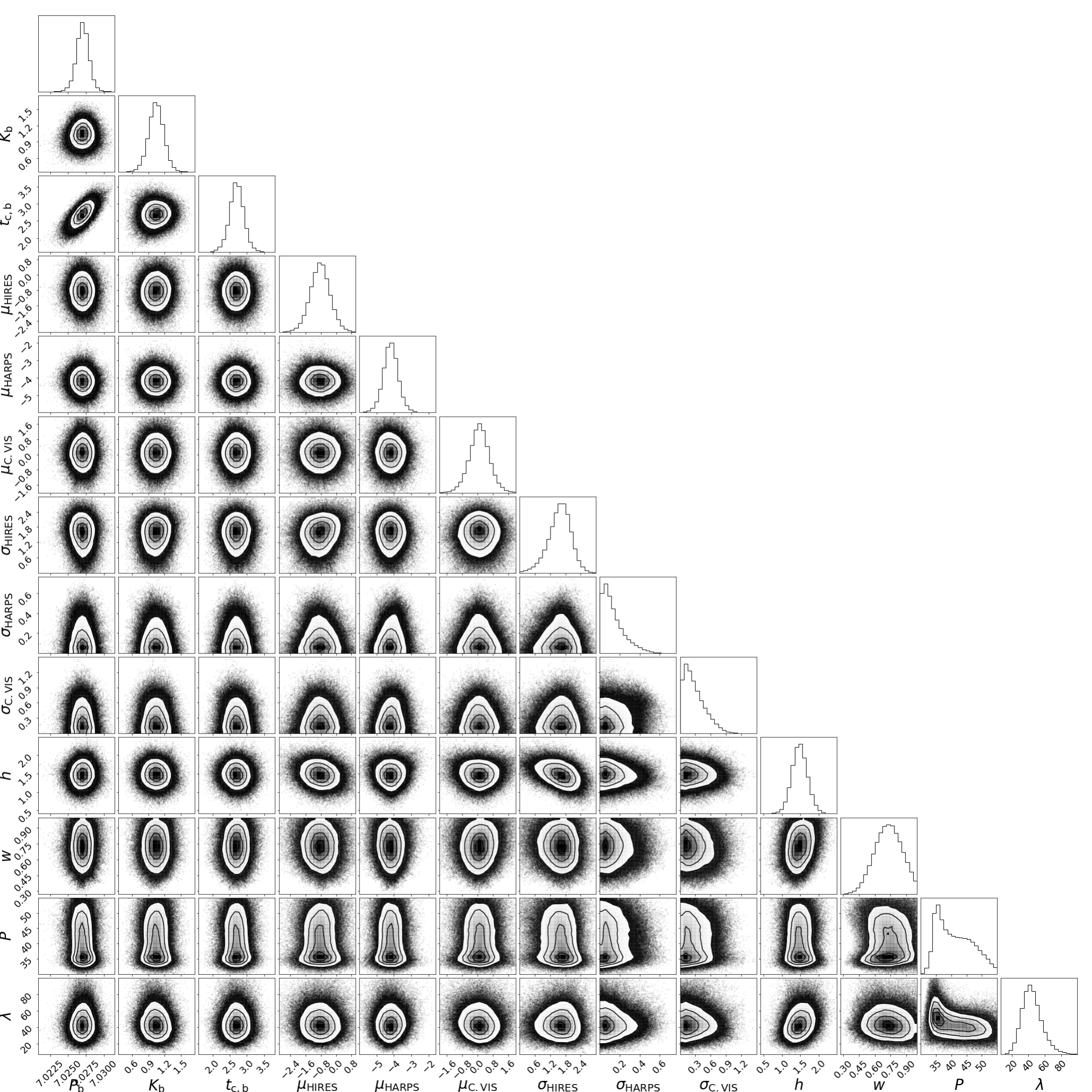}
	\caption[]{MCMC posterior distribution of the HIRES, HARPS, and CARMENES VIS RVs for Gl393 for a model including one circular planetary orbit (parameters $K_b$, $P_b$, and $T_{\rm c,b}$), a Gaussian Process using the {\tt george}-code and a quasi-periodic kernel (hyper-parameters $h$, $w$, $P$, and $\lambda$), offsets ($\mu_{\rm HIRES}$, $\mu_{\rm HARPS}$, and $\mu_{\rm C.VIS}$), and additional jitters ($\sigma_{\rm HIRES}$, $\sigma_{\rm HARPS}$, and $\sigma_{\rm C.VIS}$). Only solutions with $\ln{\mathcal L}>-716$ are shown.}
	\label{fig:cornerGl393} 
\end{figure} 		

\clearpage

\section{RV data}

\noindent{\smallskip}

\begin{table}[h!]
	\caption{Radial velocities of G264-012.$^a$}
	\label{tab:RVs G264} 
	\begin{tabular}{@{}lrrl@{}}
		\hline \hline \noalign{\smallskip}
	\multicolumn{1}{c}{BJD} & RV [m\,s$^{-1}$] & $\sigma_{\rm{RV}}$ [m\,s$^{-1}$] & Instrument \\	
		\noalign{\smallskip} 
		\hline 
		\noalign{\smallskip}
2457559.62513	&	-1.20	&	1.20	&	CARM-VIS   \\
2457597.54554	&	-1.01	&	1.40	&	CARM-VIS   \\
2457607.52688	&	-7.97	&	1.72	&	CARM-VIS   \\
2457613.49483	&	-4.27	&	1.68	&	CARM-VIS   \\
2457614.47815	&	-6.42	&	1.48	&	CARM-VIS   \\
2457618.51114	&	0.47	&	1.61	&	CARM-VIS   \\
2457619.51990	&	-6.91	&	1.55	&	CARM-VIS   \\
2457621.49704	&	-11.12	&	1.55	&	CARM-VIS   \\
2457622.37542	&	-5.41	&	1.51	&	CARM-VIS   \\
2457625.46228	&	0.79	&	1.40	&	CARM-VIS   \\
...             &   ...     &   ...     &  ...         \\
		\noalign{\smallskip}
		\hline 
	\end{tabular} 
    \tablefoot{$^a$The full table is provided at CDS. We show here the first ten rows as a guidance.}
\end{table}

\begin{table}[h!]
	\caption{Radial velocities of Gl 393.$^a$}
	\label{tab:RVs Gl393}  
	\begin{tabular}{@{}lrrl@{}}
		\hline \hline \noalign{\smallskip}
	\multicolumn{1}{c}{BJD} & RV [m\,s$^{-1}$] & $\sigma_{\rm{RV}}$ [m\,s$^{-1}$] & Instrument \\	
		\noalign{\smallskip} 
		\hline 
		\noalign{\smallskip}
2450607.77613	&	-11.63	&	1.65	&	HIRES	\\
2450807.12895	&	4.43	&	1.76	&	HIRES	\\
2450837.91365	&	-8.07	&	2.04	&	HIRES	\\
2450839.02573	&	-1.67	&	1.74	&	HIRES	\\
2450861.94499	&	-3.49	&	1.96	&	HIRES	\\
2450955.86970	&	-1.28	&	2.02	&	HIRES	\\
2451172.05304	&	1.26	&	1.84	&	HIRES	\\
2451227.97882	&	-6.25	&	1.90	&	HIRES	\\
2451342.79714	&	-10.30	&	1.89	&	HIRES	\\
2451544.16493	&	-3.30	&	2.20	&	HIRES	\\
...             &   ...     &   ...     &  ...      \\
		\noalign{\smallskip}
		\hline 
	\end{tabular} 
    \tablefoot{$^a$ The full table is provided at CDS. We show here the first ten rows as a guidance.
    }
\end{table}

\end{appendix}

\end{document}